\documentclass[12pt]{article}
\pdfoutput=1
\usepackage{epsfig}
\usepackage{amsfonts}
\usepackage{amssymb}
\usepackage{xcolor}
\usepackage{mathrsfs}
\usepackage{amsmath}
\usepackage{upgreek}
\usepackage{graphicx}
\usepackage{cite}
\usepackage{calligra}
\usepackage{caption}
\usepackage{subcaption}
\usepackage{cite}
\usepackage{calligra}
\usepackage{hyperref}
\usepackage{tikz-feynman}
\newcommand{\nc}{\newcommand}
\nc{\beq}{\begin{equation}} \nc{\eeq}{\end{equation}}
\nc{\beqa}{\begin{eqnarray}} \nc{\eeqa}{\end{eqnarray}}
\nc{\ba}{\begin{array}} \nc{\ea}{\end{array}}

\def\hc{\ensuremath{\mathrm{h.c.}}}

\newcommand{\tr}{{\text{\,tr}}}
\begin{document}
\begin{center}

{\bf \LARGE Chiral effective potential in $4D,\,\mathcal{N}=1$ supersymmetric gauge  theories} \vspace{1.0cm}

{\bf \large I.L. Buchbinder$^{1,2,3}$, R.M. Iakhibbaev$^{1}$, A.I. Mukhaeva$^{1}$, \\[0.3cm] D.I. Kazakov$^{1}$ and  D.M. Tolkachev$^{1,4}$}

\vspace{0.5cm}
{\it $^1$Bogoliubov Laboratory of Theoretical Physics, Joint Institute for Nuclear Research,
  6, Joliot Curie, 141980 Dubna, Russia\\
$^2$ Center of Theoretical Physics, Tomsk State Pedagogical University,
634041 Tomsk, Russia \\
$^3$ National Research Tomsk State University,
Lenin Av. 36, 634050 Tomsk, Russia \\and \\
$^4$Stepanov Institute of Physics,
68, Nezavisimosti Ave., 220072, Minsk, Belarus}
\vspace{0.5cm}

\abstract{We calculate the chiral effective superpotential in $4D$\, $\mathcal{N}=1,\, SU(N)$  super Yang-Mills theory coupled to chiral matter in one- and two-loop approximations. It is found that the one-loop contribution to the chiral effective potential is always finite and is expressed in terms of a specific triangle integral. The two-loop contributions generated by purely chiral vertices turned out to be finite as well. The chiral effective potential stipulated by supergraphs with gauge superfield subgraphs is finite for the supergraphs with no divergent subgraphs. In the case of the finite $\mathcal{N}=2$ SYM theory, the two-loop chiral contributions to the effective action are significanlty simplified. The leading large $N$ behavior of the chiral effective superpotential in finite $\mathcal{N}=2$ super-Yang-Mills models with $SU(N)$ gauge symmetry is studied and it is shown that the exact form in the coupling constant of the chiral effective superpotential can be found.}
\end{center}

\text{\footnotesize{ $^a$buchbinder@theor.jinr.ru, $^b$yaxibbaev@jinr.ru, 
$^c$mukhaeva@theor.jinr.ru,}} 

\text{\footnotesize{$^d$kazakovd@theor.jinr.ru, $^e$den3.1415@gmail.com,
}}

\section{Introduction}
The study of the structure of the effective action in supersymmetric gauge theories is a rich and fascinating topic. These theories often exhibit, at first glance, unexpected properties, the
understanding of which may shed light on the general structure of the quantum effective
action (see, e.g., \cite{Terning}). In cases where supersymmetric theories admit formulation in
terms of $\mathcal{N}=1$ or $\mathcal{N}=2$ superspace (see, e.g., \cite{Superspace:1983nr,BK:book,Galperin:2001seg}), unexpected properties of these theories in the quantum domain can be explained simply enough on the basis of the corresponding
non-renormalization theorems. In particular, in the $\mathcal{N}=1$ supersymmetric theories,
the renormalization theorem states that any loop contribution to the effective action is expressed
by integrals over the full superspace but not over the chiral or antichiral subspace (see, e.g., \cite{Superspace:1983nr,BK:book}). This circumstance imposes strict restrictions on the structure of ultraviolet (UV) divergences, reducing the number of possible counterterms, and sometimes allows one to construct completely finite quantum field models.

In this paper, we  discuss the structure of  low-energy contributions to the
effective action in the four-dimensional $SU(N)$,  $\mathcal{N}=1$ supersymmetric Yang-Mills theory
coupled to chiral-antichiral matter. In the theory under consideration, the leading low-energy
corrections should be the chiral effective potential that depends only on the chiral superfield and
the corresponding conjugate antichiral effective potential. Given the non-renormalization
theorem, one might think that these contributions to the effective action are forbidden.
Note, however, that this theorem says nothing about whether the contributions to the
effective action are local or not. Therefore, the non-local contributions to the effective action
are entirely acceptable. Certainly, the UV divergent contributions should be space-time
local, as for non-local contributions, there is no reason why these contributions should be
prohibited. Namely, the non-local contributions to the effective action are responsible for the chiral effective superpotential. It is known that chiral quantum corrections can indeed contribute to the one-particle irreducible (1PI) effective action in a number of ways \cite{West:1997vm,Argurio:2003ym,ShifmanVainsteinHolo}\footnote{Note that in the Wilson effective action, the chiral contributions are indeed forbidden due to the
direct cutoff of both high-energy and low-energy modes\cite{Argurio:2003ym,Bilal:2007ne}.}.

The mechanism for generating the chiral effective potential is based on the following identity
\begin{equation}
\label{identity}
\int d^8 z ~ u(\Phi) \left(-\frac{D^2}{4 \square}\right) v(\Phi)=\int d^6z ~u(\Phi) v(\Phi).
\end{equation}
where the supercoordinates are $z=(x, \theta^\alpha,\theta_{\dot{\alpha}})$ and integration measures are $d^8z=d^4x d^2\theta d^2\bar{\theta}$ and $d^6z=d^4x d^2\theta$ and $u$ and $v$ are some regular functions.
The integral in the left-hand side of \eqref{identity} is written as an integral over the full superspace,
but it is transformed into an integral over the chiral subspace due to the non-local operator
$-\frac{D^2}{4 \square}$. It is interesting to discuss how such an identity can originate in perturbation theory.
Since the divergent contributions are local in space and time, the above non-locality
can only arise as a finite contribution. The only source of the denominator can be the
massless propagator. Therefore, it can be argued that identity \eqref{identity} will only arise in
a massless theory and will correspond to the last integration in the $L$-loop supergraph when
all subgraphs are already renormalized.

The quantum corrections to the classical chiral potential in the component approach were
first discussed in \cite{West:1990rm} and calculated in \cite{JackJW:1990pd}. A manifest superfield calculation was given in \cite{BKP:1994xq}
in the context of the Wess-Zumino model. It was shown that chiral corrections appear at
the two-loop level and are represented by the non-planar vertex-type supergraph. This quantum correction is finite, which is consistent with the discussion after \eqref{identity}. Surprisingly, the two-loop finite chiral quantum correction also arises in the general chiral non-renormalizable
$\mathcal{N}=1$ model with arbitrary K\"ahler and chiral classical superpotentials \cite{Buchbinder:1999ui}. The results of calculations of chiral quantum corrections were further developed in \cite{Buchbinder:2025dgu} where the three-loop chiral effective potential in the Wess-Zumino model was obtained. It is worth pointing out that the chiral quantum corrections violate the holomorphy
of the original action, so these quantum corrections can affect the classical equations
of motion of the original model \cite{West:1991qt}. These observations led to a discussion
on supersymmetry breaking and differences between the Wilsonian and standard 1PI action
approaches (see e.g, \cite{Argurio:2003ym,ShifmanVainsteinHolo,Seiberg:1993vc,Poppitz:1996na}).
The calculation of the quantum correction to K\"ahler potential and other non-chiral quantum corrections was initiated by the work  \cite{BKY:1994iw} and discussed in detail in \cite{Pickering:1996he,Grisaru:1996ve,Banin:2006db,BuchbinderCvetic,Martin:2024qmi,Kuzenko:2014ypa}.

Chiral quantum corrections in the effective action in gauge supersymmetric theories were
discussed in \cite{West:1991qt} and \cite{BuchbinderCvetic}  only at the one-loop level. At the two-loop order, some specific
properties of the chiral effective potentials were considered in \cite{Mauri:2006uw} in the context of studying
chiral rings within the $\beta$-deformed $\mathcal{N}=1$ supersymmetric Yang-Mills model. However,
in general, direct calculations of chiral contributions to effective potentials at higher
loops remain unexplored for gauge supersymmetric models, despite some interest in this
problem, noted for example in the works \cite{ShifmanVainsteinHolo,West:1991qt}.

The problem of higher loop corrections to the chiral effective superpotential is the main
topic of the presented work. We calculate the effective chiral superpotential in higher
loops for $\mathcal{N}=1$ gauge supersymmetric theory with two conjugated chiral superfields
in the fundamental representation and one chiral superfield in the adjoint representation within
the background superfield splitting formalism. This model essentialy allows one to extend the
supersymmetry up to the $\mathcal{N}=2$ Yang-Mills model. The main attention is paid to the two-loop corrections to
the effective chiral potential in the non-abelian $\mathcal{N}=1$ super Yang-Mills theory with the
interaction of chiral superfields in the (anti)fundamental representation of the $SU(N)$ group
and in the adjoint one.

A general analysis of the possibility of obtaining chiral quantum corrections in  $\mathcal{N}=1$ supersymmetric theories containing the  chiral superfields was given in \cite{Buchbinder:1999ui} (see also the review \cite{Petrov:2001pce}) in the context of the general chiral superfield model. It was shown that these corrections arise from supergraphs where the number of $D^2$ factors is one more than the number of $\bar{D}^2$ factors. As we will see, this rule work perfectly in the theories containing gauge superfields as well.

The paper is organized as follows. In Section \ref{sec:nonminN=1} we describe the super Yang-Mills action and its effective action in the framework of the background field splitting. We derive Feynman rules for this theory and discuss the main strategy of obtaining the chiral effective superpotential. In Section \ref{sec:1loop}  we carry out one-loop calculations and show that the chiral 
quantum correction indeed exists and is finite. Section \ref{sec:div_2loop} is devoted to the calculation
of the two-loop effective superpotential where we obtain the divergent corrections
appearing due to divergent one-loop two-point subgraphs.  Also in  Section  \ref{sec:fin_2loop}  we derive
the chiral effective potential in the finite $\mathcal{N}=2$ SYM theory and discuss general properties of
this kind of potential. In the last Section \ref{sec:largeN} we discuss higher loop behaviour of the  $\mathcal{N}=1$ supersymmetric $SU(N)$ Yang-Mills action and its large $N$-color limit. The results are briefly discussed in Conclusion \ref{sec:Concl}.

\section{$\mathcal{N}=1$ super-Yang-Mills model}\label{sec:nonminN=1}

\subsection{Effective action and background splitting}
In this section, we consider the $\mathcal{N}=1$  supersymmetric Yang-Mills theory
with the gauge group $SU(N)$ which describes the interaction of vector gauge fields $V$, matter
chiral superfields $\Phi$, $\Psi_1$ and $\Psi_2$ and their corresponding conjugates:
\beq
\begin{gathered}
\mathcal{S}_c=\int d^4x ~\mathcal{L}=\tr\int d^8z~ e^{-2g V} \bar{\Phi} e^{2g V}\Phi+ \frac{1}{2}\tr\int d^6z~\mathcal{W}^2+\\ +\int d^8z~(\bar{\Psi}_1 e^{ 2gV} \Psi_1+\bar{\Psi}_2 e^{-2gV} \Psi_2)+\lambda \int d^6 z~  \Psi_1 \Phi \Psi_2+\hc,
\label{eq:n2SYM}
\end{gathered}
\eeq
where $W_\alpha=-\frac{1}{8}\bar{D}^2(e^{-2gV}D_\alpha e^{2gV})$. The real vector superfield $V$ and the real chiral superfield $\Phi$ are in the adjoint representation of $SU(N)$, whereas $\Psi_1$ transforms in the fundamental representation  and $\Psi_2$ in the antifundamental representation of $SU(N)$ and having $N_f$ copies, $g$ and $\lambda$ are coupling constants.
Actions like this extending supersymmetric quantum chromodynamics are of interest as independent models in connection with electric-magnetic dualities  and studies of physics on infrared behaviour of a wide class of supersymmetric models (see e.g. \cite{Kutasov:1995ve,Kutasov:1995ss,Brodie:1996vx}). Here, however, we find it to be convenient for our purposes in calculating the chiral effective superpotential.

The model \eqref{eq:n2SYM} explicitly contains (anti-)chiral vertices, so that non-holomorphic
contributions to the effective potential can be obtained as well. For the sake of convenience, the classical tree-level potential is denoted as follows:
\begin{equation}
    W_{tree}=\lambda \int d^6 z~  \Psi_1 \Phi \Psi_2.
\end{equation}

The action \eqref{eq:n2SYM} is invariant under supergauge transformations
\begin{equation}
\begin{gathered}
   e^{2gV'}\rightarrow e^{ig\bar{\Lambda}} e^{2gV}e^{-ig\Lambda},
\end{gathered}
\end{equation}
where $\bar{D}_{\dot{\alpha}}\Lambda=0$.  Chiral superfield is also gauge transformed as
\begin{align}
  \Phi'&=e^{ig\Lambda}\Phi e^{-ig\Lambda}, \quad
  \bar\Phi'=e^{ig\bar\Lambda}\bar\Phi e^{-ig\bar\Lambda},\nonumber\\
  \Psi'_1&=e^{ig\Lambda}\Psi_1, \quad  \Psi'_2=e^{-ig\Lambda}\Psi_2,
    \end{align}
where $\Lambda=\Lambda_a T^a$ with the corresponding adjoint gauge generators $(T^a)_{bc}=f^{abc}$.  One has the following identites for the $SU(N)$ group generators $[t_a,t_b]=i (T^c)_{ab} t_c$, $(t^a)^i_j (t^a)^j_k=C_F\delta^i_k$, $T_{abc}T^{c'}_{ab}=C_A\delta^{c'}_c$, $\text{tr}[t_a t_b]=T_A\delta_{ab}$ where $C_F=\frac{N^2-1}{2N}$, $C_A=N$ and $T_F=1/2$ is the Dynkin index of the fundamental representation of the gauge group $SU(N)$. 

The unitarity and nonabelian gauge invariance conditions generate important terms. The first, gauge-fixing term is as follows:
\begin{equation}
\mathcal{S}_{GF}=-\frac{1}{16\xi}\text{tr}\int d^8z~ D^2 V \bar{D}^2 V.
\end{equation}
We work in the Feynman gauge $\xi=1$ since this gauge cancels IR-singularities at the one-loop level \cite{Superspace:1983nr}. A convenience of choosing this gauge is that the one-loop two-point subgraphs with vector closed lines are identically zero. Note that the gauge fixing term does not include the superfields $\Phi,\phi$ and $\Psi,\psi$.  
And due to the unitarity condition we have to include the Faddeev-Popov ghost term which is given by
\beq
\mathcal{S}_{FP} =\text{tr}\int d^8z~\left[ \bar{c}'c - c' \bar{c}
+\frac{1}{2}(c'+\bar{c}')[V, c+ \bar{c}] + \cdots \right],
\eeq
Note that $c,c'$  are  (anti)chiral ghost superfields and are given in the adjoint representation of the gauge group. 
So the full action of the model can be written as
\begin{equation}
\mathcal{S}_0=\mathcal{S}_c+\mathcal{S}_{GF}+\mathcal{S}_{FP}.
\label{eq:action}
\end{equation}

All subsequent considerations are based on the study of the structure of the quantum effective action for  $\mathcal{N} = 1$ supersymmetric Yang-Mills teories constructed in the standard way (see, e.g., \cite{BK:book}). The calculations are carried out within the framework of the loop expansion of the effective action that is based on the background quantum splitting of the initial fields of the action~\eqref{eq:action}. Since our aim is the chiral effective porential, the background fields are introduced only for the chiral superfields $\Phi, \Psi_1, \Psi_2$. The gauge superfield $V$ can be only quantum. The corresponding background quantum splitting has the form 
\begin{equation}
\Phi \rightarrow \Phi+\sqrt{\hbar}\phi, ~\Psi_I \rightarrow \Psi_I+\sqrt{\hbar}\psi_I, \label{chiralsplitting}
\end{equation}
where $\phi$ and $\psi_I$ ($I=1,2$) are the quantum superfields and both terms in \eqref{chiralsplitting} transform under the corresponding gauge transformations.

As usual, the quadratic part of the action in the quantum fields $\phi$ and $\psi_I$  defines the propagators, and the other part defines the vertices. The corresponding quadratic part of the action~\eqref{eq:action} after background quantum splitting \eqref{chiralsplitting} is written as follows:
\begin{align}
\mathcal{S}^{(2)}&= \tr\int d^8 z ~\left(\bar{\phi} \phi  -2g^2 \bar{\Phi} [v, \phi]+2g^2 \bar{\phi} [v, \Phi] \right) \nonumber\\ \nonumber
&\int d^8z~(\bar{\psi}_1\psi_1-g\bar{\Psi}_1 v \psi_1-g\bar{\psi}_1 v \Psi_1) \nonumber \\&\int d^8z~(\bar{\psi}_2\psi_2+g\bar{\Psi}_2 v \psi_2+g\bar{\psi}_2 v \Psi_2) \nonumber \\
    &-\lambda\int d^6z~ \left(\psi_1 \Phi \psi_{2}+\Psi_1 \phi \psi_{2}+\psi_1 \phi \Psi_{2}+ \hc\right) \nonumber \\
    &+\tr\int d^8z~\left(-\frac{1}{2} v\hat{\square}v +cc^\prime+c^\prime c\right). \label{Squadr}
\end{align}
where the quantum gauge superfield $V$ is redefiend as $v$ for uniformity with $\phi, \psi_1, \psi_2$. The rest part of the action $\mathcal{S}_c+\mathcal{S}_{GF}+\mathcal{S}_{FP}$  determines interaction vertices of the quantum fields.
The details of the discussion on deriving Feynman rules from the effective supersymmetric action for all sectors of the considered model can be found, e.g., in \cite{BK:book,Superspace:1983nr}.

In the case where the background chiral superfields $\Phi, \Psi_I$ and $\bar{\Phi}, \bar{\Psi}_I$ satisfy the %supersymmetric 
conditions
\begin{equation}\label{conditions}
\partial_a\Phi=\partial_a \Psi_I = \partial_a\bar{\Phi}=\partial_a\bar{\Psi}_I = 0,
\end{equation}
%(i.e. they shoud be independent on 4-coordinates)
one can get an effective potential from the effective action (note, however, that for covariant spinor deritives there hold
$\bar{D}_{\dot{\alpha}}\Phi=0, ~\bar{D}_{\dot{\alpha}}\Psi_I=0,  ~D^{\alpha}\bar{\Phi}=0, ~D^{\alpha}\bar{\Psi}_I=0
$). If the background gauge superfield is zero, the superfield effective potential can be found in the following form \cite{BKY:1994iw}
\beq
\label{sep}
\Gamma[\Phi,\Psi_i|\bar{\Phi},\bar{\Psi}_i]= \int d^8z~ \left(\textbf{K} +
\textbf{A}\right)
+ \left(\int d^6z~ \textbf{W} + \hc\right),
\eeq
where $\textbf{K}(\Phi,\Psi_I|\bar{\Phi},\bar{\Psi}_I)$ is the Kähler effective potential (it depends only on the combination of the corresponding fields and their conjugates), $\textbf{A}(\Phi,D\Phi,\Psi_I,D\Psi_I|\bar{\Phi},\bar{D}\bar{\Phi},\bar{\Psi}_I,\bar{D}\bar{\Psi}_I)$ is the effective superpotential of auxiliary fields and  $\textbf{W}(\Phi,\Psi_I)$ is the chiral effective superpotential. The latter is the main focus of this paper.
In the framework of loop expansion, the superfield
effective potential can be written as a series in the number of loops
\beq
\label{loopexp1}
\Gamma[\Phi,\Psi_I|\bar{\Phi},\bar{\Psi}_I]=\sum_{L=1}^\infty\hbar^L \Gamma^{(L)}[\Phi,\Psi_I|\bar{\Phi},\bar{\Psi}_I].
\eeq
Therefore, each of the effective potentials $\textbf{K},\, \textbf{A}$ and $\textbf{W}$ can also be written in the form \eqref{loopexp1}  \cite{BK:book,BKY:1993,Pickering:1996he}, e.g.,
\beq
\label{loopexp}
\textbf{W}[\Phi,\Psi_I]=\sum_{L=1}^\infty\hbar^L \textbf{W}^{(L)}[\Phi,\Psi_I].
\eeq
Based on expression \eqref{Squadr}, one can obtain the supersymmetric Green function for each of the quantized superfields. Then, they can be re-expressed in such a way that the corresponding propagators do not contain background antichiral fields $\bar{\Phi},\bar{\Psi}_I$. The product of these Green functions as a result of the calculations does not contain antichiral fields, and the integration yield an expression corresponding to the quantum correction to the classical chiral superpotential. The general procedure for obtaining the chiral effective superpotential using the improved supergraph technique is discussed in detail in Ref. \cite{BK:book}.

The possibility of obtaining chiral loop corrections to the classical potential was
analyzed in Refs. \cite{BKY:1993} where it was shown that such corrections are possible under the condition
\beq
 n_{D^2}+1=n_{\bar{D}^2}, \label{cond}
\eeq
where $n_D$ is a number of covariant chiral derivatives in a generic connected 1PI supergraph. Note that for pure chiral theories, contributions to the chiral effective potential are only possible beginning with the two-loop level \cite{BKY:1993}, whereas interaction with the gauge field allows one to satisfy the condition \eqref{cond} even at the one-loop level. A list of Feynman rules needed in the calculation of this part of the effective action for the considered super Yang-Mills model is given in Appendix \ref{sec:Feyn_diagr}. In the following sections, based on \eqref{cond} and the obtained Feynman rules, we construct and calculate chiral quantum corrections to the superpotential up to two loops.

\section{One-loop chiral integral}\label{sec:1loop}

We now turn to the study of possible chiral corrections to the effective action. It turns out that in this theory~\eqref{eq:n2SYM}, firstly, we have a one-loop correction to the chiral effective potential, and, secondly, it is finite.

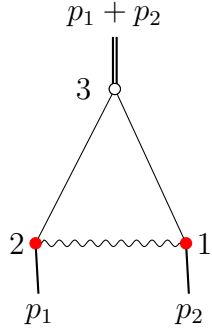
\begin{figure}[htb]
    \centering
\begin{subfigure}{.3\textwidth}
	\centering
	\begin{tikzpicture}[
		scale=1,
		mydot/.style={
			circle,
			fill=white,
			draw,
			outer sep=0pt,
			inner sep=1.5pt
		}
		]
		\begin{feynman}
			\vertex  at (0,2) (a1){\(p_1+p_2\)};
			\node[below=1cm of a1, mydot] (a2);
			\vertex at (0,1) (a2) ;
			\vertex at (-1, -1) (d); \vertex[below left=0.0cm of d, red, dot] (d){};
			\vertex at (1, -1) (e); \vertex[below left=0.0cm of e, red,dot] (e){};
			\vertex at (-1,-2) (f) {\(p_1\)};
			\vertex at (1, -2) (g) {\(p_2\)};
			\diagram*{
				(a2) --[double,thick] (a1),
				(d) -- [plain] (a2),
				(a2) -- [plain] (e),
				(e) -- [boson] (d) ,
				(f) -- [plain,thick] (d),
				(g) -- [plain,thick] (e),
			};
			\vertex [left=0.4em of a2] {\(3\)};
			\vertex [left=0.6em of d] {\(2\)};
			\vertex [right=0.6em of e] {\(1\)};
			\node[below=1cm of a1, mydot] (a2);
		\end{feynman}
	\end{tikzpicture}
\end{subfigure}%
    \caption{Feynman one-loop supergraph contributing to the chiral superpotential. Red dots denote vertex $g$, white dots denote chiral vertices $\lambda$. External thick lines denote classical background fields, thin lines denote quantum lines. Double lines denote the chiral field in the adjoint representation, while thin plain lines represent the chiral fields in the (anti)fundamental representation and curved line is the vector superpropagator. }
    \label{Fig:oneloop}
\end{figure}

As noted above, for an arbitrary graph that contributes to the chiral effective potential, the number of covariant derivatives $D^2$ must contain one more than the number of anticovariant derivatives  $\bar{D}^2$. There is only one chiral graph at the one-loop level which can be constructed under the obtained Feynman rules (see Appendix \ref{sec:Feyn_diagr}). This integral has the following form (shown in Fig. \ref{Fig:oneloop}):
\beq
\begin{gathered}
   \textbf{W}^{(1)}= \lim_{p_1,p_2 \rightarrow 0}\lambda g^2(2C_F-C_A) \int \prod_{l=1}^3 d^8 z_l ~\Phi(z_1)\Psi_{1}(z_2) \Psi_{2}(z_3)\\\left\{ \frac{1}{\square_2}\delta_{1,2} \frac{D^2_1 \bar{D}_3}{16{\square_1}}\delta_{1,3}\frac{D^2_2}{4}\delta_{2,3}   \frac{1}{\square_2}\right\}.
\end{gathered}
\eeq
Note that in calculations, we first set external momenta to be nonzero. We note that the momentum integral itself is proportional to $1/p^2$. But the numerator of the integral due to \eqref{identity} is proportional to $p^2$. Thus, after calculating the momentum integral, we take the limit of external momenta tending to zero. This leads us to a finite expression which contributes exactly to the chiral effective potential. The described procedure extracts the needed contribution from the effective action and corresponds to the definition of the effective potential under the condition \eqref{conditions}.
To proceed here, it is nessesary to integrate over the free delta function and use the standard relation for covariant derivatives $D$, see the rules in Appendix \ref{sec:AppA}) and then make use of \eqref{identity}.
After contracting group indices and performing the corresponding $D$-algebra, we make the assumption that the superfields are slowly changing
$$\Psi_1(y_1,\theta) \Psi_1(y_2,\theta) \Phi(x,\theta)\simeq [\Psi_1 \Phi \Psi_2] (x,\theta),$$
where $y_1,y_2,x$ are the 4-coordinates, i.e., all chiral superfields are defined at one point in the full superspace.
Thus, integration results in
\beq
\label{one-loop}
 \textbf{W}^{(1)}=  \frac{\hbar}{(4\pi)^2} g^2 (2C_F-C_A)\Upsilon^{(1)} W_{tree}.
\eeq
Note that the expression $2C_F-C_A$ can not be vanishing for the $SU(N)$ group. 
There $\Upsilon^{(1)}$ is the reduced Davydychev-Usyukina triangle integral \cite{Usyukina:1992jd}:
\beq
\Upsilon^{(1)}=\lim_{p_1,p_2 \rightarrow 0} \int d^4 q \frac{(p_1+p_2)^2}{q^2 (q-p_1)^2 (q_1+p_2)^2}=\int_0^1 d\tau \frac{2 \log(\tau)}{\tau^2-\tau+1}.
\label{eq:Ups1}
\eeq
We take into account possible cyclic permutation of external fields because the gauge propagator can be inserted in every side of the triangle graph under consideration, but as the superfields $\Phi$ and $V$ are in the adjoint representation interacting through the structure constant, the contributions cancel each other.
The result (\ref{one-loop}) is in  agreement with the calculations carried out in Ref. \cite{West:1991qt} where more general model was studied. The model considered here is more specific in comparison with  \cite{West:1991qt}. Nevertheless, it is general enough and allows one to reduce the group structure to Casimir operators. In principle, one can represent the integral (\ref{eq:Ups1}) in the form of finite combination of polygamma functions or multiple zeta-functions  of rational argument (see Appendix \ref{sec:AppB}), but we would instead  prefer to present the result (\ref{one-loop}) through integral (\ref{eq:Ups1}) since in the latter in the higher loop case we will obtain generalizations just such an integral.

Relation (\ref{one-loop}) is the final result for the one-loop contribution to the chiral effective potential. It is important to emphasize that the obtained contribution is automatically finite. Next, we proceed to the analysis of higher loops.

\section{Two-loop supergraph calculations}\label{sec:2loop}

In this section, we are going to consider two-loop quantum corrections to the chiral potential.
The first part \ref{sec:2l_chiral} consists in calculating the purely chiral contribution. The second \ref{sec:div_2loop} and third \ref{sec:fin_2loop} parts contain divergent and finite contributions to the effective potential, respectively. At the end \ref{sec:fin_res}, we arrive at the full two-loop result for the superpotential.

\subsection{Two-loop purely  chiral superfield graph}\label{sec:2l_chiral}

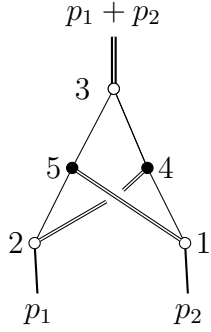
\begin{figure}
    \centering
\begin{subfigure}{.3\textwidth}
	\centering
	\begin{tikzpicture}[
		scale=1,
		mydot/.style={
			circle,
			fill=white,
			draw,
			outer sep=0pt,
			inner sep=1.5pt
		}
		]
		\begin{feynman}
			\vertex  at (0,2) (a1){\(p_1+p_2\)};
			\node[below=1cm of a1, mydot] (a2);
			\vertex at (0,1) (a2) ;
			\vertex at (-0.5, 0) (b); \vertex[below left=0.0cm of b, black,dot] (b){};
			\vertex at (0.5, 0) (c);  \vertex[below left=0.0cm of c, black,dot] (c){};
			\vertex at (0.125, -0.35) (cc);
			\vertex at (-1, -1) (d); \vertex[below left=0.0cm of d, empty  dot] (d){};
			\vertex at (-0.1, -0.5) (dd);
			\vertex at (1, -1) (e); \vertex[below left=0.0cm of e, empty  dot] (e){};
			\vertex at (-1,-2) (f) {\(p_1\)};
			\vertex at (1, -2) (g) {\(p_2\)};
			\diagram*{
				(a2) --[double,thick] (a1),
				(b) -- [plain] (a2),
				(a2) -- [plain] (c),
				(a2) -- [plain] (c),
				(e)-- [double] (b) ,
				(d) -- [double] (dd),
				(cc) -- [double] (c),
				(c) -- [plain] (e),
				(d) -- [plain] (b),
				(f) -- [plain,thick] (d),
				(g) -- [plain,thick] (e),
			};
			\vertex [left=0.4em of a2] {\(3\)};
			\vertex [left=0.6em of b] {\(5\)};
			\vertex [right=0.6em of c] {\(4\)};
			\vertex [left=0.6em of d] {\(2\)};
			\vertex [right=0.6em of e] {\(1\)};
			\node[below=1cm of a1, mydot] (a2);
		\end{feynman}
	\end{tikzpicture}
\end{subfigure}%    
    \caption{Feynman two-loop chiral supergraph contributing to the non-holomorphic superpotential. Here white dots denote chiral vertices $\lambda$ and black dots denote anti-chiral vertices $\bar{\lambda}$. }
    \label{Fig:twoloopchiral}
\end{figure}

In the absence of gauge interaction, contributions to the chiral effective superpotential start at two-loop order, since one-loop graphs cannot be assembled from only chiral massless internal lines  \cite{BKP:1994xq} (there are no $\langle \phi_i \phi_j\rangle$-propagators). This non-planar two-loop diagram is depicted in Fig. \ref{Fig:twoloopchiral} and can explicitly be written as
\beq
\begin{gathered}
\textbf{W}^{(2)}=\lim_{p_1,p_2\rightarrow0}2|\lambda|^4(C_A-C_F)(C_A-2C_F)  \int \prod_{l=1}^5 d^8 z_l~ \lambda\Psi_1(z_3)\Phi(z_4)\Psi_2(z_5) \nonumber\\\left\{ \frac{1}{\square_1}\delta_{1,3} \frac{D^2_2 \bar{D}^2_3}{16\square_2} \delta_{3,2}   \frac{1}{16 \square_2} \delta_{2,4} \frac{D^2_1\bar{D}^2_4}{16\square_1} \delta_{1,4} \frac{D^2_1\bar{D}^2_5}{16\square_1} \delta_{1,5}\frac{D^2_2}{4\square_2}\delta_{2,5} \right\}.
\end{gathered}\label{twoloopDint}
\eeq
The integral can be evaluated after $D$-algebra reduction. In this case, the $D$-algebra routine can be performed using \texttt{SusyMath.m}\cite{Ferrari:2007sc} or by hand and integration of scalar master-integrals can be performed with the help of Feynman parametrization.
Despite the fact that this diagram is not divergent, in some approaches to calculating Feynman integrals, it is much simpler to apply methods that use dimensional regularization $d=4-2\epsilon$ \cite{Buchbinder:2025dgu}.

The scalar master-integral after $D$-algebra is given by an integral  in momentum space
\beq
\begin{gathered}
I^{(2)}=\lim_{p_{1,2}\rightarrow0}\int \frac{d^4 q_1}{(4\pi)^4} \frac{d^4 q_2}{(4\pi)^4}  \frac{q_1^2 p_1^2+q_2^2 p_2^2-2 p_1 p_2 (q_1 q_2)}{q_1^2q_2^2 (q_1+q_2)^2 (q_1-p_1)^2 (q_2-p_2)^2 (q_1+q_2-p_1-p_2)^2}.
\label{twoloopmaster}
\end{gathered}
\eeq
The last expression can be calculated with the help of integration by parts and the uniqueness relation (see details in \cite{Buchbinder:2025dgu,Kazakov:1983dyk,Grozin:2012xi}). The result is as follows
\beq
\textbf{W}^{(2)}=\frac{\hbar^2}{(4\pi)^4}12(C_A-C_F)(C_A-2C_F)|\lambda|^4 \zeta(3) \times W_{tree},
\label{eq:2l_chiral}
\eeq
where $\zeta(n)$ is the Riemann zeta-function (hereafter we omit $\hbar$-factor for the sake of compactness, we can restore this dependence after exchanging $g^2 \rightarrow g^2\hbar, ~|\lambda|^2 \rightarrow |\lambda|^2 \hbar$) and the numerical prefactor is the symmetry factor of the diagram. This contribution, as can be seen, is finite but suppressed by the number of colors $N$, as the corresponding diagram is non-planar. This two-loop finite contribution is purely chiral and contains only chiral interactions $\lambda$ and $\bar\lambda$. This diagram was first calculated in component form in Ref. \cite{JackJW:1990pd} and in superfield form in Ref. \cite{BKP:1994xq,BKY:1994iw}, as noted in the Introduction. In the following subsection we consider two-loop diagrams containing gauge superfield propagators.

\subsection{Divergent two-loop supergraphs with gauge interaction}\label{sec:div_2loop}

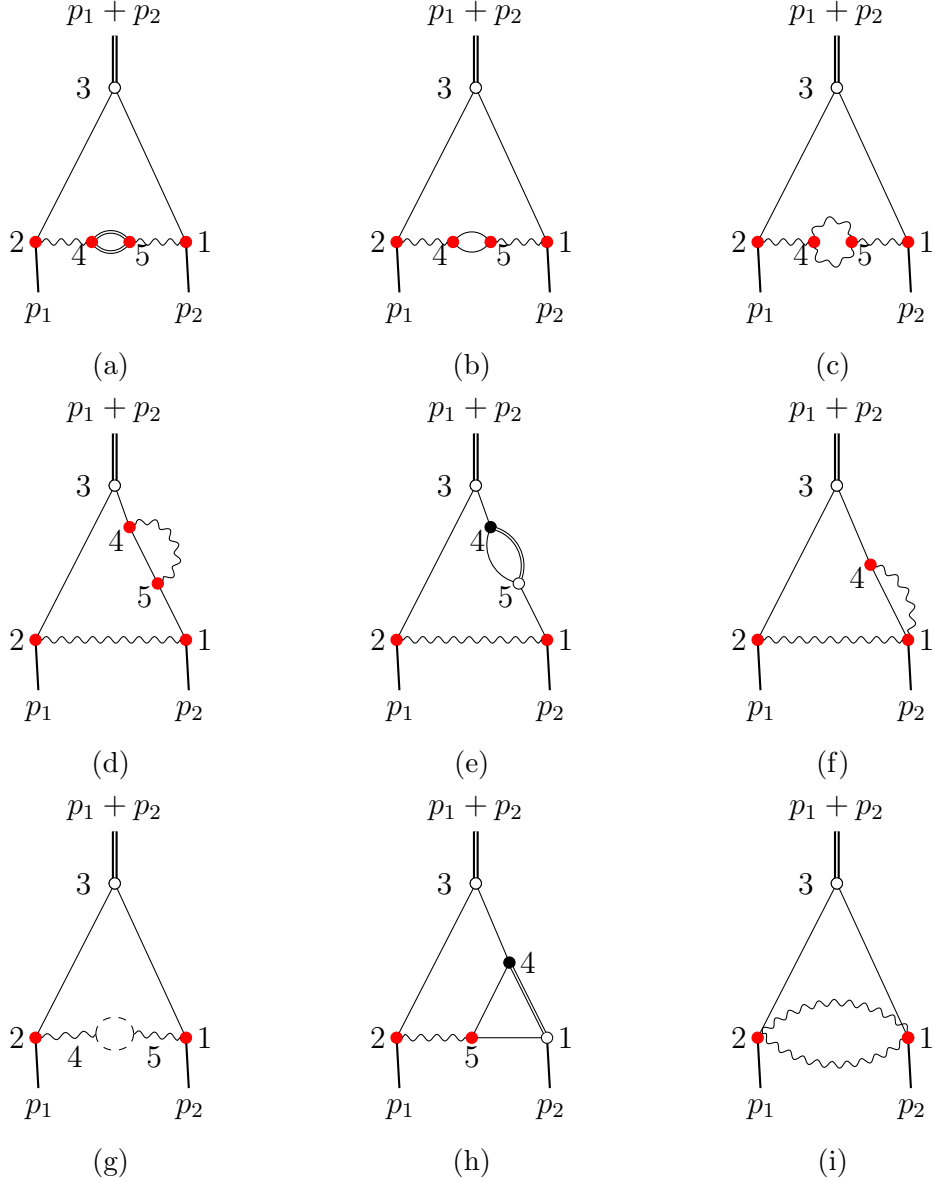
\begin{figure}[]
        \centering        
        	\begin{subfigure}{.3\textwidth}%a
        	\centering
        	\begin{tikzpicture}[
        		scale=1,
        		mydot/.style={
        			circle,
        			fill=white,
        			draw,
        			outer sep=0pt,
        			inner sep=1.5pt
        		}
        		]
        		\begin{feynman}[layered layout, horizontal=b to c]
        			\vertex  at (0,2) (a1){\(p_1+p_2\)};
        			\node[below=1cm of a1, mydot] (a2);
        			\vertex at (0,1) (a2) ;
        			\vertex at (-1, -1) (b); \vertex[below left=0.0cm of b,red, dot] (b){};
        			\vertex at (1, -1) (c);  \vertex[below left=0.0cm of c,red, dot] (c){};
        			\vertex at (-0.25,-1) (d); \vertex[below left=0.0cm of d,  red,dot] (d){};
        			\vertex at (0.25, -1) (e); \vertex[below left=0.0cm of e,red, dot] (e){};
        			\vertex at (-1,-2) (f) {\(p_1\)};
        			\vertex at (1, -2) (g) {\(p_2\)};
        			\diagram*[layered layout,horizontal'=d to e] {
        				(a2) --[double,thick] (a1),
        				(b) -- [plain] (a2),
        				(a2) -- [plain] (c),
        				(b) -- [boson] (d)
        				-- [double, quarter left] (e)
        				-- [double, quarter left] (d),
        				(e) -- [boson] (c),
        				(g) -- [plain,thick] (c),
        				(f) -- [plain,thick] (b),
        			};
        			\vertex [left=0.4em of a2] {\(3\)};
        			\vertex [left=0.6em of b] {\(2\)};
        			\vertex [right=0.6em of c] {\(1\)};
        			\vertex [below left=0.6em of d] {\(4\)};
        			\vertex [below right=0.6em of e] {\(5\)};
        			\node[below=1cm of a1, mydot] (a2);
        		\end{feynman}
        	\end{tikzpicture}
        	\caption{}
        \end{subfigure}%
        \begin{subfigure}{.3\textwidth}%b
        	\centering
        	\begin{tikzpicture}[
        		scale=1,
        		mydot/.style={
        			circle,
        			fill=white,
        			draw,
        			outer sep=0pt,
        			inner sep=1.5pt
        		}
        		]
        		\begin{feynman}[layered layout, horizontal=b to c]
        			\vertex  at (0,2) (a1){\(p_1+p_2\)};
        			\node[below=1cm of a1, mydot] (a2);
        			\vertex at (0,1) (a2) ;
        			\vertex at (-1, -1) (b); \vertex[below left=0.0cm of b,red, dot] (b){};
        			\vertex at (1, -1) (c);  \vertex[below left=0.0cm of c,red, dot] (c){};
        			\vertex at (-0.25,-1) (d); \vertex[below left=0.0cm of d,  red,dot] (d){};
        			\vertex at (0.25, -1) (e); \vertex[below left=0.0cm of e,red, dot] (e){};
        			\vertex at (-1,-2) (f) {\(p_1\)};
        			\vertex at (1, -2) (g) {\(p_2\)};
        			\diagram*[layered layout,horizontal'=d to e] {
        				(a2) --[double,thick] (a1),
        				(b) -- [plain] (a2),
        				(a2) -- [plain] (c),
        				(b) -- [boson] (d)
        				-- [plain, quarter left] (e)
        				-- [plain, quarter left] (d),
        				(e) -- [boson] (c),
        				(g) -- [plain,thick] (c),
        				(f) -- [plain,thick] (b),
        			};
        			\vertex [left=0.4em of a2] {\(3\)};
        			\vertex [left=0.6em of b] {\(2\)};
        			\vertex [right=0.6em of c] {\(1\)};
        			\vertex [below left=0.6em of d] {\(4\)};
        			\vertex [below right=0.6em of e] {\(5\)};
        			\node[below=1cm of a1, mydot] (a2);
        		\end{feynman}
        	\end{tikzpicture}
        	\caption{}
        \end{subfigure}%
        \begin{subfigure}{.3\textwidth}%c
        	\centering
        	\begin{tikzpicture}[
        		scale=1,
        		mydot/.style={
        			circle,
        			fill=white,
        			draw,
        			outer sep=0pt,
        			inner sep=1.5pt
        		}
        		]
        		\begin{feynman}[layered layout, horizontal=b to c]
        			\vertex  at (0,2) (a1){\(p_1+p_2\)};
        			\node[below=1cm of a1, mydot] (a2);
        			\vertex at (0,1) (a2) ;
        			\vertex at (-1, -1) (b); \vertex[below left=0.0cm of b,red, dot] (b){};
        			\vertex at (1, -1) (c);  \vertex[below left=0.0cm of c, red,dot] (c){};
        			\vertex at (-0.25,-1) (d); \vertex[below left=0.0cm of d,  red,dot] (d){};
        			\vertex at (0.25, -1) (e); \vertex[below left=0.0cm of e, red,dot] (e){};
        			\vertex at (-1,-2) (f) {\(p_1\)};
        			\vertex at (1, -2) (g) {\(p_2\)};
        			\diagram*[layered layout,horizontal'=d to e] {
        				(a2) --[double,thick] (a1),
        				(b) -- [plain] (a2),
        				(a2) -- [plain] (c),
        				(b) -- [boson] (d)
        				-- [boson, half left] (e)
        				-- [boson, half left,] (d),
        				(e) -- [boson] (c),
        				(g) -- [plain,thick] (c),
        				(f) -- [plain,thick] (b),
        			};
        			\vertex [left=0.4em of a2] {\(3\)};
        			\vertex [left=0.6em of b] {\(2\)};
        			\vertex [right=0.6em of c] {\(1\)};
        			\vertex [below left=0.6em of d] {\(4\)};
        			\vertex [below right=0.6em of e] {\(5\)};
        			\node[below=1cm of a1, mydot] (a2);
        		\end{feynman}
        	\end{tikzpicture}
        	\caption{}
        \end{subfigure}\\
        \begin{subfigure}{.3\textwidth}%d
        	\centering
        	\begin{tikzpicture}[
        		scale=1,
        		mydot/.style={
        			circle,
        			fill=white,
        			draw,
        			outer sep=0pt,
        			inner sep=1.5pt
        		}
        		]
        		\begin{feynman}
        			\vertex  at (0,2) (a1){\(p_1+p_2\)};
        			\node[below=1cm of a1, mydot] (a2);
        			\vertex at (0,1) (a2) ;
        			\vertex at (-1, -1) (b); \vertex[below left=0.0cm of b, red,dot] (b){};
        			\vertex at (1, -1) (c);  \vertex[below left=0.0cm of c, red,dot] (c){};
        			\vertex at (0.25, 0.5) (d); \vertex[below left=0.0cm of d, red, dot] (d){};
        			\vertex at (0.625,-0.25) (e); \vertex[below left=0.0cm of e,red, dot] (e){};
        			\vertex at (-1,-2) (f) {\(p_1\)};
        			\vertex at (1, -2) (g) {\(p_2\)};
        			\diagram* {
        				(a2) --[double,thick] (a1),
        				(b) -- [plain] (a2),
        				(a2) -- [plain] (d),
        				(d) -- [plain] (e),
        				(e) -- [plain] (c),
        				(b) -- [boson] (c),
        				(d) -- [boson, half left] (e),
        				(g) -- [plain,thick] (c),
        				(f) -- [plain,thick] (b),
        			};
        			\vertex [left=0.4em of a2] {\(3\)};
        			\vertex [left=0.6em of b] {\(2\)};
        			\vertex [right=0.6em of c] {\(1\)};
        			\vertex [below left=0.6em of d] {\(4\)};
        			\vertex [below left=0.6em of e] {\(5\)};
        			\node[below=1cm of a1, mydot] (a2);
        		\end{feynman}
        	\end{tikzpicture}
        	\caption{}
        \end{subfigure}%
        \begin{subfigure}{.3\textwidth}%e
        	\centering
        	\begin{tikzpicture}[
        		scale=1,
        		mydot/.style={
        			circle,
        			fill=white,
        			draw,
        			outer sep=0pt,
        			inner sep=1.5pt
        		}
        		]
        		\begin{feynman}
        			\vertex  at (0,2) (a1){\(p_1+p_2\)};
        			\node[below=1cm of a1, mydot] (a2);
        			\vertex at (0,1) (a2) ;
        			\vertex at (-1, -1) (b); \vertex[below left=0.0cm of b,red, dot] (b){};
        			\vertex at (1, -1) (c);  \vertex[below left=0.0cm of c, red,dot] (c){};
        			\vertex at (0.25, 0.5) (d); \vertex[below left=0.0cm of d, dot] (d){};
        			\vertex at (0.625,-0.25) (e); \vertex[below left=0.0cm of e,  empty dot] (e){};
        			\vertex at (-1,-2) (f) {\(p_1\)};
        			\vertex at (1, -2) (g) {\(p_2\)};
        			\diagram* {
        				(a2) --[double,thick] (a1),
        				(b) -- [plain] (a2),
        				(b) -- [boson] (c),
        				(a2) -- [plain] (d)
        				-- [double, quarter left] (e)
        				-- [plain, quarter left] (d),
        				(e) -- [plain] (c),
        				(g) -- [plain,thick] (c),
        				(f) -- [plain,thick] (b),
        			};
        			\vertex [left=0.4em of a2] {\(3\)};
        			\vertex [left=0.6em of b] {\(2\)};
        			\vertex [right=0.6em of c] {\(1\)};
        			\vertex [below left=0.6em of d] {\(4\)};
        			\vertex [below left=0.6em of e] {\(5\)};
        			\node[below=1cm of a1, mydot] (a2);
        		\end{feynman}
        	\end{tikzpicture}
        	\caption{}
        \end{subfigure}%
        \begin{subfigure}{.3\textwidth}%f
        	\centering
        	\begin{tikzpicture}[
        		scale=1,
        		mydot/.style={
        			circle,
        			fill=white,
        			draw,
        			outer sep=0pt,
        			inner sep=1.5pt
        		}
        		]
        		\begin{feynman}
        			\vertex  at (0,2) (a1){\(p_1+p_2\)};
        			\node[below=1cm of a1, mydot] (a2);
        			\vertex at (0,1) (a2) ;
        			\vertex at (-1, -1) (b); \vertex[below left=0.0cm of b, red,dot] (b){};
        			\vertex at (1, -1) (c);  \vertex[below left=0.0cm of c, red,dot] (c){};
        			\vertex at ( 0.5,0) (d); \vertex[below left=0.0cm of d, red, dot] (d){};
        			\vertex at (0.25,0.5) (x); \vertex at (-0.5,1) (xx);
        			\vertex at (-1,-2) (f) {\(p_1\)};
        			\vertex at (1, -2) (g) {\(p_2\)};
        			\diagram* {
        				(a2) --[double,thick] (a1),
        				(b) -- [plain] (a2),
        				(a2) -- [plain] (d),
        				(b) -- [boson] (c),
        				(d) -- [boson, quarter left] (c),
        				(c) -- [plain] (d),
        				(g) -- [plain,thick] (c),
        				(f) -- [plain,thick] (b)
        			};
        			\vertex [left=0.4em of a2] {\(3\)};
        			\vertex [left=0.6em of b] {\(2\)};
        			\vertex [right=0.6em of c] {\(1\)};
        			\vertex [below left=0.6em of d] {\(4\)};
        			\node[below=1cm of a1, mydot] (a2);
        		\end{feynman}
        	\end{tikzpicture}
        	\caption{}
        \end{subfigure}\\
        \begin{subfigure}{.3\textwidth}%g
        	\centering
        	\begin{tikzpicture}[
        		scale=1,
        		mydot/.style={
        			circle,
        			fill=white,
        			draw,
        			outer sep=0pt,
        			inner sep=1.5pt
        		}
        		]
        		\begin{feynman}[layered layout, horizontal=b to c]
        			\vertex  at (0,2) (a1){\(p_1+p_2\)};
        			\node[below=1cm of a1, mydot] (a2);
        			\vertex at (0,1) (a2) ;
        			\vertex at (-1, -1) (b); \vertex[below left=0.0cm of b,red, dot] (b){};
        			\vertex at (1, -1) (c);  \vertex[below left=0.0cm of c, red,dot] (c){};
        			\vertex at (-0.25,-1) (d); %\vertex[below left=0.0cm of d,  red,dot] (d){};
        			\vertex at (0.25, -1) (e); %\vertex[below left=0.0cm of e, red,dot] (e){};
        			\vertex at (-1,-2) (f) {\(p_1\)};
        			\vertex at (1, -2) (g) {\(p_2\)};
        			\diagram*[layered layout,horizontal'=d to e] {
        				(a2) --[double,thick] (a1),
        				(b) -- [plain] (a2),
        				(a2) -- [plain] (c),
        				(b) -- [boson] (d)
        				-- [scalar, half left] (e)
        				-- [scalar, half left,] (d),
        				(e) -- [boson] (c),
        				(g) -- [plain,thick] (c),
        				(f) -- [plain,thick] (b),
        			};
        			\vertex [left=0.4em of a2] {\(3\)};
        			\vertex [left=0.6em of b] {\(2\)};
        			\vertex [right=0.6em of c] {\(1\)};
        			\vertex [below left=0.1em of d] {\(4\)};
        			\vertex [below right=0.1em of e] {\(5\)};
        			\node[below=1cm of a1, mydot] (a2);
        		\end{feynman}
        	\end{tikzpicture}
        	\caption{}
        \end{subfigure}%
        \begin{subfigure}{.3\textwidth}%h
        	\centering
        	\begin{tikzpicture}[
        		scale=1,
        		mydot/.style={
        			circle,
        			fill=white,
        			draw,
        			outer sep=0pt,
        			inner sep=1.5pt
        		}
        		]
        		\begin{feynman}
        			\vertex  at (0,2) (a1){\(p_1+p_2\)};
        			\node[below=1cm of a1, mydot] (a2);
        			\vertex at (0,1) (a2) ;
        			\vertex at (0, -1) (b); \vertex[below left=0.0cm of b, red,dot] (b){};
        			\vertex at (0.5, 0) (c);  \vertex[below left=0.0cm of c, dot] (c){};
        			\vertex at (-1, -1) (d); \vertex[below left=0.0cm of d, red, dot] (d){};
        			\vertex at (1, -1) (e); \vertex[below left=0.0cm of e,empty dot] (e){};
        			\vertex at (-1,-2) (f) {\(p_1\)};
        			\vertex at (1, -2) (g) {\(p_2\)};
        			\diagram*{
        				(a2) --[double,thick] (a1),
        				(d) -- [plain] (a2),
        				(a2) -- [plain] (c),
        				%(a2) -- [plain] (c),
        				(c) -- [double] (e),
        				(e) -- [plain] (b),
        				(b)-- [boson] (d) ,
        				(c) -- [plain] (b),
        				%(e) -- [boson, momentum=\(k\)] (d) ,
        				(f) -- [plain,thick] (d),
        				(g) -- [plain,thick] (e),
        			};
        			\vertex [left=0.4em of a2] {\(3\)};
        			\vertex [below=0.6em of b] {\(5\)};
        			\vertex [right=0.6em of c] {\(4\)};
        			\vertex [left=0.6em of d] {\(2\)};
        			\vertex [right=0.6em of e] {\(1\)};
        			\node[below=1cm of a1, mydot] (a2);
        		\end{feynman}
        	\end{tikzpicture}
        	\caption{}
        \end{subfigure}%
        \begin{subfigure}{.3\textwidth}%i
        	\centering
        	\begin{tikzpicture}[
        		scale=1,
        		mydot/.style={
        			circle,
        			fill=white,
        			draw,
        			outer sep=0pt,
        			inner sep=1.5pt
        		}
        		]
        		\begin{feynman}[layered layout, horizontal=b to c]
        			\vertex  at (0,2) (a1){\(p_1+p_2\)};
        			\node[below=1cm of a1, mydot] (a2);
        			\vertex at (0,1) (a2) ;
        			\vertex at (-1, -1) (b); \vertex[below left=0.0cm of b, red,dot] (b){};
        			\vertex at (1, -1) (c);  \vertex[below left=0.0cm of c, red,dot] (c){};
        			\vertex at (-1,-1) (d);
        			\vertex at (1, -1) (e);
        			\vertex at (-1,-2) (f) {\(p_1\)};
        			\vertex at (1, -2) (g) {\(p_2\)};
        			\diagram*[layered layout,horizontal'=d to e] {
        				(a2) --[double,thick] (a1),
        				(b) -- [plain] (a2),
        				(a2) -- [plain] (c),
        				(b) -- [boson] (d)
        				-- [boson, quarter left] (e)
        				-- [boson, quarter left] (d),
        				(e) -- [boson] (c),
        				(g) -- [plain,thick] (c),
        				(f) -- [plain,thick] (b),
        			};
        			\vertex [left=0.4em of a2] {\(3\)};
        			\vertex [left=0.6em of b] {\(2\)};
        			\vertex [right=0.6em of c] {\(1\)};
        			\vertex[below left=0.0cm of c, red,dot] (c){};
        			\node[below=1cm of a1, mydot] (a2);
        		\end{feynman}
        	\end{tikzpicture}
        	\caption{}
        \end{subfigure}\\
	\caption{Feynman two-loop divergent diagrams contributing to the chiral superpotential. Straight lines are chiral superpropagators, curved line is a vector superpropagator. Red dots denote chiral-vector vertex $\sim g$, white dots denote chiral vertices $\lambda$, and black dots denote anti-chiral vertices $\bar{\lambda}$. }
\label{fig:divergent}
\end{figure}

Here it is appropriate to discuss diagrams containing divergences, which is why we consider below all formally divergent diagrams appearing in perturbation theory expansion.
First, we consider two-loop diagrams with possibly UV divergent subgraphs with gauge interactions, satysfying \eqref{cond}, see Fig.\ref{fig:divergent} . Usually, two-point subgraphs at the single-loop level are divergent  \cite{WestBook}. Let us write out the integrals in more detail. The integrals with the chiral and ghost superfield loop have the following form, see Fig.\ref{fig:divergent} a), b) and g)
\beq
\begin{gathered}
\textbf{W}^{(2),A}_{div}+\textbf{W}^{(2),B}_{div}+\textbf{W}^{(2),G}_{div}=\lim_{p_1,p_2\rightarrow0} (2 N_f T_F-C_A) (2C_F-C_A) 4g^4 \lambda \times  \\ \times \int \prod_{l=1}^5 d^8 z_l  ~ \Psi_1(z_1)\Phi(z_2)\Psi_2(z_3)  \left\{ \frac{D^2_1 \bar{D}^2_3}{16\square_1} \delta_{1,3}   \frac{D^2_2}{4\square_2} \delta_{3,2} \frac{1}{\square_4}\delta_{2,4} \frac{\bar{D}^2_4 D^2_5}{16\square_4} \delta_{4,5}  \frac{\bar{D}^2_5 D^2_4}{16\square_5} \delta_{5,4}\frac{1}{\square_1}\delta_{5,1} \right\}.
\end{gathered}\label{D2a}
\eeq

Integrals with the gauge superfield loop in Fig.\ref{fig:divergent} c) have the following form:
\beq
\begin{gathered}
\textbf{W}^{(2),C}_{div}\sim \lim_{p_1,p_2\rightarrow0}4g^4 \lambda \int \prod_{l=1}^5 d^8 z_l ~\Psi_1(z_1)\Phi(z_2)\Psi_2(z_3) \left\{ \frac{D^2_1 \bar{D}^2_3}{16\square_1} \delta_{1,3}  \right. \\ \left. \frac{D^2_2}{4\square_2} \delta_{3,2}\frac{1}{\square_4}\delta_{2,4}\frac{1}{\square_4}\delta_{4,5}  \frac{1}{\square_5} \delta_{5,4}\frac{1}{\square_1}\delta_{5,1} \right\}=0,
\end{gathered}\label{D2b}
\eeq
and Fig.\ref{fig:divergent} i) can automatically be found as
\beq
\begin{gathered}
\textbf{W}^{(2),I}_{div}=0.
\end{gathered}\label{D2c}
\eeq
These integrals~\eqref{D2b}-\eqref{D2c} are identically zero exactly due to the Feynman-Fermi gauge, as mentioned above.

\begin{figure}
    \centering
\begin{minipage}[h]{0.3\linewidth}
	\vspace{0pt}
        \includegraphics[width=\textwidth]{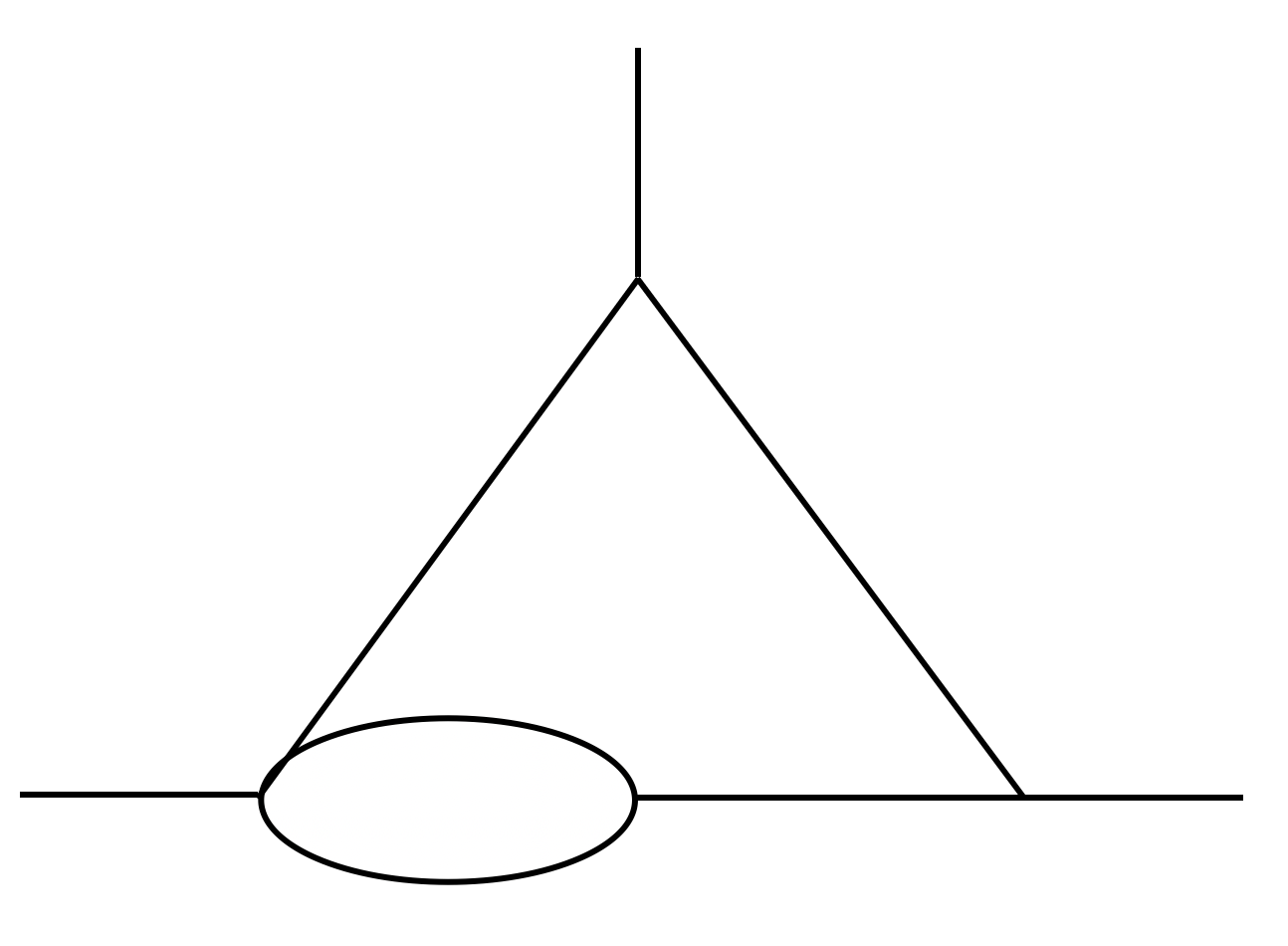}
         \end{minipage}
        \caption{All divergent diagrams are equivalent to the triangle scalar master integral with one-loop subgraph inserted in the internal propagator.}
\label{fig:master_div}
\end{figure}

Integration for triangle diagrams with one-loop insertions results in the well-known formula of the one-loop of the $\mathcal{N}=1$ supersymmetric Yang-Mills theory multiplied by the triangle integral shown in Fig.(\ref{fig:master_div}) \footnote{The integrals are computed within the framework of dimensional regularisation within the standard technique after $D$-algebra routine.}
    \beq   \textbf{W}^{(2)}_{div,1}=\text{A+B+G}=4g^4\left(2 N_f T_{F}- C_A\right) (2C_F-C_A)\times   J^{(1)}_{1,1}\Upsilon^{(1)} \times W_{tree},
    \label{eq:J11}
    \eeq
where $ J^{(1)}_{1,1}(k) $ is the one-loop bubble integral, see Eq.~\eqref{eq:J_11} from Appendix \ref{sec:AppB}. Its explicit form looks like
    \beq
    J^{(1)}_{1,1}(k) =\left(\frac{1}{\epsilon}+2+O(\epsilon^{1})\right)(k^2/\mu^2)^{-\epsilon},
    \eeq
and  $\Upsilon^{(1)} $ is given by~\eqref{eq:Ups1}. The meaning of expression \eqref{eq:J11} is very simple and is related to the number of fields running along the inner two-point loop and their representations. Here the $d=4-2\epsilon$ regularization is introduced and the minimal subtraction scheme (MS) is adopted, i.e., all constant terms arising from one-loop divergent subgraphs are absorbed in the parameter of dimensional transmutation $\mu^2$.

The Feynman integrals $\textbf{W}^{(2),D}_{div}$ and $\textbf{W}^{(2),E}_{div}$ (Fig.\ref{fig:divergent} d) and e)) are integrated as
\beq
   \textbf{W}^{(2)}_{div,2}= \text{2D+2E}=2g^2 (2C_F-C_A) \left(|\lambda|^2 - g^2\right)\times J^{(1)}_{1,1}~\Upsilon^{(1)} \times W_{tree}.
    \label{eq:anom_c}
    \eeq
The integrals $\textbf{W}^{(2),F}$ and $\textbf{W}^{(2),H}$ (Fig.\ref{fig:divergent} f), h)) have the same form as $\textbf{W}^{(2),D}$ and $\textbf{W}^{(2),E}$, respectively.
Thus, the final result for the divergent part of the chiral effective superpotential at the two-loop level can be expressed as follows:
\begin{equation}
    \textbf{W}^{(2)}_{div}=\left(g^4\left\{2N_fT_F-C_A\right\}+g^2(|\lambda^2|-g^2)\right)~4(2C_F-C_A)J^{(1)}_{1,1}\Upsilon^{(1)} \times W_{tree}
    \label{N1Wdiv}
\end{equation}

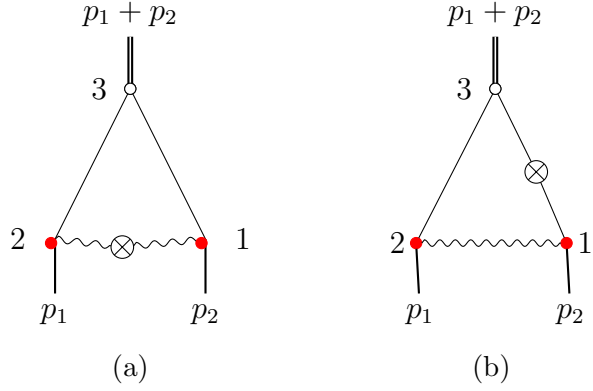
\begin{figure}
    \centering
   \begin{subfigure}{.3\textwidth}
   	\centering
   	\begin{tikzpicture}[
   		scale=1,
   		mydot/.style={
   			circle,
   			fill=white,
   			draw,
   			outer sep=0pt,
   			inner sep=1.5pt
   		}
   		]
   		\begin{feynman}
   			\vertex  at (0,2) (a1){\(p_1+p_2\)};
   			\node[below=1cm of a1, mydot] (a2);
   			\vertex at (0,1) (a2) ;
   			\vertex at (-1, -1) (d); 
   			\vertex at (1, -1) (e); 
   			\vertex at (-1.1, -1.1) (d1); 
   			\vertex at (1.01, -1.01) (e1); 
   			\vertex at (0, -1) (ee);  \vertex[below left=0.0cm of ee, crossed dot] (ee){};
   			\vertex at (-1,-2) (f) {\(p_1\)};
   			\vertex at (1, -2) (g) {\(p_2\)};
   			\diagram*{
   				(a2) --[double,thick] (a1),
   				(d) -- [plain] (a2),
   				(a2) -- [plain] (e),
   				(e) --  [boson] (ee) ,
   				(ee) --  [boson] (d) ,
   				(f) -- [plain,thick] (d),
   				(g) -- [plain,thick] (e),
   			};
   			\vertex [left=0.4em of a2] {\(3\)};
   			\vertex [left=0.6em of d] {\(2\)};
   			\vertex [right=0.6em of e] {\(1\)};
   			\vertex[below left=0.0cm of d, red, dot] (d){};
   			\vertex[below left=0.0cm of e, red, dot] (e){};
   			\node[below=1cm of a1, mydot] (a2);
   		\end{feynman}
   	\end{tikzpicture}
   	\caption{}
   \end{subfigure}%
   \begin{subfigure}{.3\textwidth}
   	\centering
   	\begin{tikzpicture}[
   		scale=1,
   		mydot/.style={
   			circle,
   			fill=white,
   			draw,
   			outer sep=0pt,
   			inner sep=1.5pt
   		}
   		]
   		\begin{feynman}
   			\vertex  at (0,2) (a1){\(p_1+p_2\)};
   			\node[below=1cm of a1, mydot] (a2);
   			\vertex at (0,1) (a2) ;
   			\vertex at (-1, -1) (d); \vertex[below left=0.0cm of d, red, dot] (d){};
   			\vertex at (0.65, 0) (ee);  \vertex[below left=0.0cm of ee, crossed dot] (ee){};
   			\vertex at (1, -1) (e); \vertex[below left=0.0cm of e, red,dot] (e){};
   			\vertex at (-1,-2) (f) {\(p_1\)};
   			\vertex at (1, -2) (g) {\(p_2\)};
   			\diagram*{
   				(a2) --[double,thick] (a1),
   				(d) -- [plain] (a2),
   				(a2) -- [plain] (ee),
   				(ee)--  [crossed dot] (e),
   				(e) --  [boson] (d) ,
   				(f) -- [plain,thick] (d),
   				(g) -- [plain,thick] (e),
   			};
   			\vertex [left=0.4em of a2] {\(3\)};
   			\vertex [left=0.6em of d] {\(2\)};
   			\vertex [right=0.6em of e] {\(1\)};
   			\node[below=1cm of a1, mydot] (a2);
   		\end{feynman}
   	\end{tikzpicture}
   	\caption{}
   \end{subfigure}%
        \caption{Diagrams with counterterms inserted in (a) gauge superfield propagator and in (b) chiral superfield line.}
\label{fig:counterterms}
\end{figure}

There are some comments on the perturbative behaviour of the model in order. The last expression \eqref{eq:anom_c} leads to a non-trivial anomalous dimension in the theory under consideration. However, in the case of requiring the restoration of additional hidden supersymmetry, which implies $|\lambda|^2=g^2$, this contribution vanishes\footnote{One can state that this condition is a consequence of non-renormalization theorem for $\mathcal{N}=2$ SYM models \cite{Buchbinder:2014wra}}. It is known \cite{Buchbinder:2014wra}, the perturbative behavior of this extended supersymmetric theory at all orders will be determined solely by the one-loop divergence \eqref{eq:J_11} leading to the following beta function of $\mathcal{N}=2$ SYM theory \cite{Koh:1983ir,Howe:1983wj}
\begin{equation}
    \beta(g)=\frac{2g^3}{(4\pi)^2} (2N_fT_F-C_A).
\end{equation}
Therefore, in the $\mathcal{N}=2$ SYM model the following condition takes place \cite{Koh:1983ir,Howe:1983wj}
\begin{equation}
    \sum_i  m_i T_F(R_i)=C_A
\end{equation}
the corresponding quantum field theory will be ultraviolet finite. Here $R_i$ is some representation of the gauge group and $m_i$ is a number of hypermupltiplets. The details of discussion of the finiteness condition can be found in, e.g. \cite{Koh:1983ir,West:1983jm}. In the case of a theory with extended supersymmetry and the considered gauge group $SU(N)$, the $N_f=N$ fundamental representations are required to satisfy the finiteness condition, as can be seen from our calculations\cite{BK:book,WestBook,Howe:1983wj}. In other cases, the obtained divergences must be removed by introducing diagrams with counterterms, which are shown on Fig.\ref{fig:counterterms}, leading to renormalization of superfields and couplings as in any renormalizable model \cite{WestBook,Parkes:1984dh}.

Thus, in this subsection it is shown that the divergent quantum corrections to the chiral superpotential with the mentioned one-loop subgraphs can consistently be calculated in agreement with the results of previous calculations.

\subsection{Finite two-loop supergraphs with gauge interaction}\label{sec:fin_2loop}

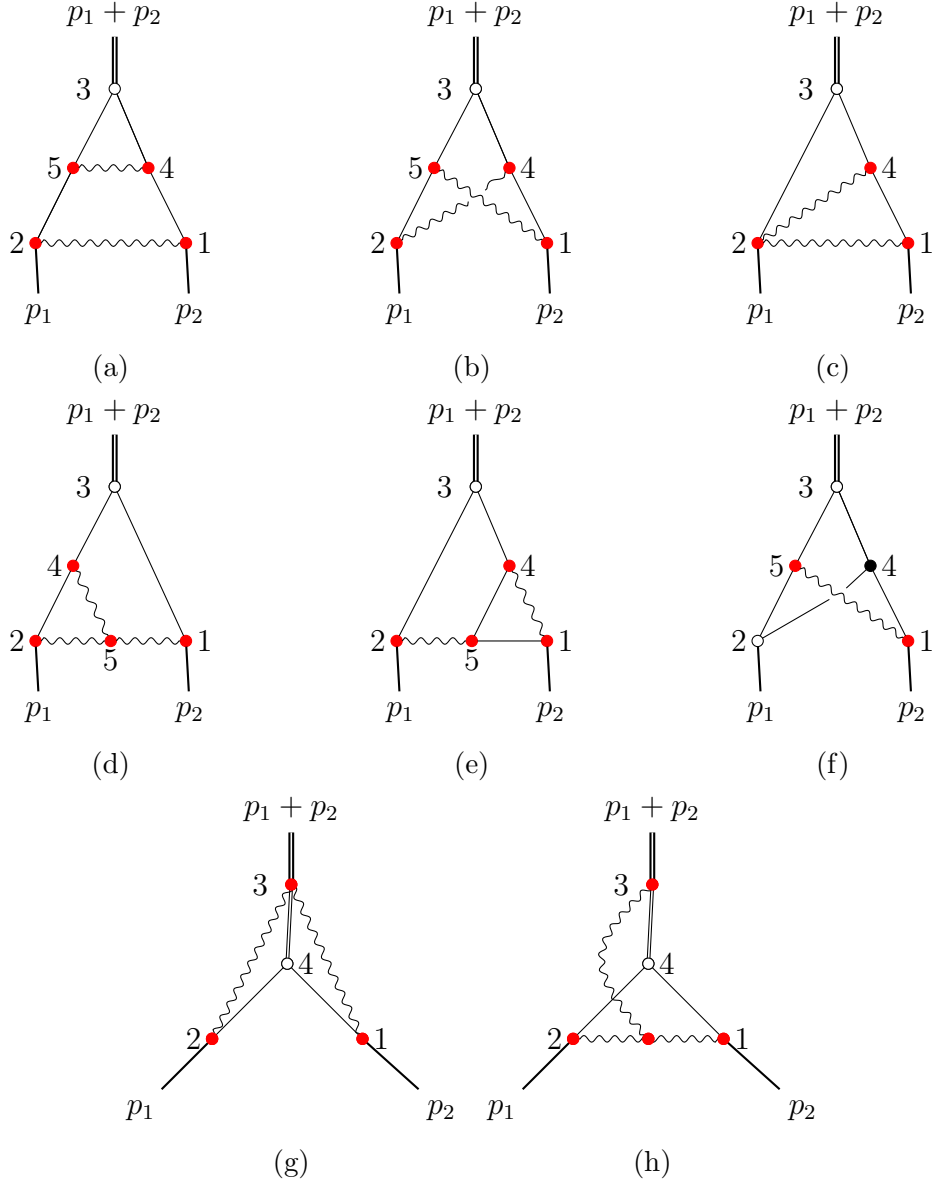
\begin{figure}[h!]
\centering
\begin{subfigure}{0.3\textwidth}
	\centering
	\begin{tikzpicture}[
		scale=1,
		mydot/.style={
			circle,
			fill=white,
			draw,
			outer sep=0pt,
			inner sep=1.5pt
		}
		]
		\begin{feynman}
			\vertex  at (0,2) (a1){\(p_1+p_2\)};
			\vertex at (0,1) (a2) ;
			\vertex at (-0.5, 0) (b); \vertex[below left=0.0cm of b, red,dot] (b){};
			\vertex at (0.5, 0) (c);  \vertex[below left=0.0cm of c,red, dot] (c){};
			\vertex at (-1, -1) (d); \vertex[below left=0.0cm of d, red, dot] (d){};
			\vertex at (1, -1) (e); \vertex[below left=0.0cm of e, red,dot] (e){};
			\vertex at (-1,-2) (f) {\(p_1\)};
			\vertex at (1, -2) (g) {\(p_2\)};
			\diagram*{
				(a2) --[double,thick] (a1),
				(b) -- [plain] (a2),
				(a2) -- [plain] (c),
				(a2) -- [plain] (c),
				(c)-- [boson] (b) ,
				(d) -- [plain] (b),
				(c) -- [plain] (e),
				(d) -- [plain] (b),
				(e) -- [boson] (d) ,
				(f) -- [plain,thick] (d),
				(g) -- [plain,thick] (e),
			};
			\vertex [left=0.4em of a2] {\(3\)};
			\vertex [left=0.6em of b] {\(5\)};
			\vertex [right=0.6em of c] {\(4\)};
			\vertex [left=0.6em of d] {\(2\)};
			\vertex [right=0.6em of e] {\(1\)};
			\node[below=1cm of a1, mydot] (a2);
		\end{feynman}
	\end{tikzpicture}
	\caption{}
\end{subfigure}%
\begin{subfigure}{.3\textwidth}
	\centering
	\begin{tikzpicture}[
		scale=1,
		mydot/.style={
			circle,
			fill=white,
			draw,
			outer sep=0pt,
			inner sep=1.5pt
		}
		]
		\begin{feynman}
			\vertex  at (0,2) (a1){\(p_1+p_2\)};
			\node[below=1cm of a1, mydot] (a2);
			\vertex at (0,1) (a2) ;
			\vertex at (-0.5, 0) (b); \vertex[below left=0.0cm of b, red,dot] (b){};
			\vertex at (0.5, 0) (c);  \vertex[below left=0.0cm of c, red,dot] (c){};
			\vertex at (0.125, -0.35) (cc);
			\vertex at (-1, -1) (d); \vertex[below left=0.0cm of d, red, dot] (d){};
			\vertex at (-0.1, -0.5) (dd);
			\vertex at (1, -1) (e); \vertex[below left=0.0cm of e, red,dot] (e){};
			\vertex at (-1,-2) (f) {\(p_1\)};
			\vertex at (1, -2) (g) {\(p_2\)};
			\diagram*{
				(a2) --[double,thick] (a1),
				(b) -- [plain] (a2),
				(a2) -- [plain] (c),
				(a2) -- [plain] (c),
				(e)-- [boson] (b) ,
				(d) -- [boson] (dd),
				(cc) -- [boson] (c),
				(c) -- [plain] (e),
				(d) -- [plain] (b),
				%	(e) -- [boson, momentum=\(k\)] (d) ,
				(f) -- [plain,thick] (d),
				(g) -- [plain,thick] (e),
			};
			\vertex [left=0.4em of a2] {\(3\)};
			\vertex [left=0.6em of b] {\(5\)};
			\vertex [right=0.6em of c] {\(4\)};
			\vertex [left=0.6em of d] {\(2\)};
			\vertex [right=0.6em of e] {\(1\)};
			\node[below=1cm of a1, mydot] (a2);
		\end{feynman}
	\end{tikzpicture}
	\caption{}
\end{subfigure}%
\begin{subfigure}{.3\textwidth}
	\centering
	\begin{tikzpicture}[
		scale=1,
		mydot/.style={
			circle,
			fill=white,
			draw,
			outer sep=0pt,
			inner sep=1.5pt
		}
		]
		\begin{feynman}
			\vertex  at (0,2) (a1){\(p_1+p_2\)};
			\node[below=1cm of a1, mydot] (a2);
			\vertex at (0,1) (a2) ;
			%	\vertex at (-1, 0) (b); \vertex[below left=0.0cm of b, dot] (b){};
			\vertex at (0.5, 0) (c);  \vertex[below left=0.0cm of c, red,dot] (c){};
			\vertex at (-1, -1) (d); \vertex[below left=0.0cm of d, red, dot] (d){};
			\vertex at (1, -1) (e); \vertex[below left=0.0cm of e, red,dot] (e){};
			\vertex at (-1,-2) (f) {\(p_1\)};
			\vertex at (1, -2) (g) {\(p_2\)};
			\diagram*{
				(a2) --[double,thick] (a1),
				(d) -- [plain] (a2),
				(a2) -- [plain] (c),
				%(a2) -- [plain] (c),
				(e)-- [boson] (d) ,
				(d) -- [boson] (c),
				(c) -- [plain] (e),
				%	(d) -- [plain] (b),
				%	(e) -- [boson, momentum=\(k\)] (d) ,
				(f) -- [plain,thick] (d),
				(g) -- [plain,thick] (e),
			};
			\vertex [left=0.4em of a2] {\(3\)};
			\vertex [right=0.6em of c] {\(4\)};
			\vertex [left=0.6em of d] {\(2\)};
			\vertex [right=0.6em of e] {\(1\)};
			\node[below=1cm of a1, mydot] (a2);
		\end{feynman}
	\end{tikzpicture}
	\caption{}
\end{subfigure}\\
\begin{subfigure}{.3\textwidth}
	\centering
	\begin{tikzpicture}[
		scale=1,
		mydot/.style={
			circle,
			fill=white,
			draw,
			outer sep=0pt,
			inner sep=1.5pt
		}
		]
		\begin{feynman}
			\vertex  at (0,2) (a1){\(p_1+p_2\)};
			\node[below=1cm of a1, mydot] (a2);
			\vertex at (0,1) (a2) ;
			\vertex at (-0.5, 0) (b); \vertex[below left=0.0cm of b, red,dot] (b){};
			\vertex at (0, -1) (c);  \vertex[below left=0.0cm of c,red, dot] (c){};
			\vertex at (-1, -1) (d); \vertex[below left=0.0cm of d, red, dot] (d){};
			\vertex at (1, -1) (e); \vertex[below left=0.0cm of e, red,dot] (e){};
			\vertex at (-1,-2) (f) {\(p_1\)};
			\vertex at (1, -2) (g) {\(p_2\)};
			\diagram*{
				(a2) --[double,thick] (a1),
				(b) -- [plain] (a2),
				(a2) -- [plain] (e),
				%(a2) -- [plain] (c),
				(c) -- [boson] (b),
				(e) -- [boson] (c),
				(c)-- [boson] (d) ,
				(d) -- [plain] (b),
				%(e) -- [boson, momentum=\(k\)] (d) ,
				(f) -- [plain,thick] (d),
				(g) -- [plain,thick] (e),
			};
			\vertex [left=0.4em of a2] {\(3\)};
			\vertex [left=0.6em of b] {\(4\)};
			\vertex [below=0.6em of c] {\(5\)};
			\vertex [left=0.6em of d] {\(2\)};
			\vertex [right=0.6em of e] {\(1\)};
			\node[below=1cm of a1, mydot] (a2);
			%  \vertex[below left=0.0cm of a3, mydot] (a3){};
		\end{feynman}
	\end{tikzpicture}
	\caption{}
\end{subfigure}%
\begin{subfigure}{.3\textwidth}
	\centering
	\begin{tikzpicture}[
		scale=1,
		mydot/.style={
			circle,
			fill=white,
			draw,
			outer sep=0pt,
			inner sep=1.5pt
		}
		]
		\begin{feynman}
			\vertex  at (0,2) (a1){\(p_1+p_2\)};
			\node[below=1cm of a1, mydot] (a2);
			\vertex at (0,1) (a2) ;
			\vertex at (0, -1) (b); \vertex[below left=0.0cm of b, red,dot] (b){};
			\vertex at (0.5, 0) (c);  \vertex[below left=0.0cm of c, red,dot] (c){};
			\vertex at (-1, -1) (d); \vertex[below left=0.0cm of d, red, dot] (d){};
			\vertex at (1, -1) (e); \vertex[below left=0.0cm of e,red, dot] (e){};
			\vertex at (-1,-2) (f) {\(p_1\)};
			\vertex at (1, -2) (g) {\(p_2\)};
			\diagram*{
				(a2) --[double,thick] (a1),
				(d) -- [plain] (a2),
				(a2) -- [plain] (c),
				%(a2) -- [plain] (c),
				(c) -- [boson] (e),
				(e) -- [plain] (b),
				(b)-- [boson] (d) ,
				(c) -- [plain] (b),
				%(e) -- [boson, momentum=\(k\)] (d) ,
				(f) -- [plain,thick] (d),
				(g) -- [plain,thick] (e),
			};
			\vertex [left=0.4em of a2] {\(3\)};
			\vertex [below=0.6em of b] {\(5\)};
			\vertex [right=0.6em of c] {\(4\)};
			\vertex [left=0.6em of d] {\(2\)};
			\vertex [right=0.6em of e] {\(1\)};
			\node[below=1cm of a1, mydot] (a2);
		\end{feynman}
	\end{tikzpicture}
	\caption{}
\end{subfigure}%
\begin{subfigure}{.3\textwidth}
\centering
\begin{tikzpicture}[
scale=1,
mydot/.style={
	circle,
	fill=white,
	draw,
	outer sep=0pt,
	inner sep=1.5pt
}
]
\begin{feynman}
	\vertex  at (0,2) (a1){\(p_1+p_2\)};
	\node[below=1cm of a1, mydot] (a2);
	\vertex at (0,1) (a2) ;
	\vertex at (-0.5, 0) (b); \vertex[below left=0.0cm of b, red,dot] (b){};
	\vertex at (0.5, 0) (c);  \vertex[below left=0.0cm of c,black,dot] (c){};
	\vertex at (0.125, -0.35) (cc);
	\vertex at (-1, -1) (d); \vertex[below left=0.0cm of d, empty dot] (d){};
	\vertex at (-0.1, -0.5) (dd);
	\vertex at (1, -1) (e); \vertex[below left=0.0cm of e, red,dot] (e){};
	\vertex at (-1,-2) (f) {\(p_1\)};
	\vertex at (1, -2) (g) {\(p_2\)};
	\diagram*{
		(a2) --[double,thick] (a1),
		(b) -- [plain] (a2),
		(a2) -- [plain] (c),
		(a2) -- [plain] (c),
		(e)-- [boson] (b) ,
		(d) -- [plain] (dd),
		(cc) -- [plain] (c),
		(c) -- [plain] (e),
		(d) -- [plain] (b),
		%	(e) -- [boson, momentum=\(k\)] (d) ,
		(f) -- [plain,thick] (d),
		(g) -- [plain,thick] (e),
	};
	\vertex [left=0.4em of a2] {\(3\)};
	\vertex [left=0.6em of b] {\(5\)};
	\vertex [right=0.6em of c] {\(4\)};
	\vertex [left=0.6em of d] {\(2\)};
	\vertex [right=0.6em of e] {\(1\)};
	\node[below=1cm of a1, mydot] (a2);
\end{feynman}
\end{tikzpicture}
\caption{}
\end{subfigure}\\
\begin{subfigure}{.3\textwidth}
\centering
\begin{tikzpicture}[
scale=1,
mydot/.style={
	circle,
	fill=white,
	draw,
	outer sep=0pt,
	inner sep=1.5pt
}
]
\begin{feynman}
	\vertex  at (0,2) (a1){\(p_1+p_2\)};
	\node[below=1cm of a1, red, dot] (a2);
	\vertex  at (0,0) (a3); \vertex[below left=0.0cm of a3, empty dot] (a3){};
	\vertex at (0,1) (a2) ;
	\vertex at (-1, -1) (b); \vertex[below left=0.0cm of b, red,dot] (b){};
	\vertex at (1, -1) (c);  \vertex[below left=0.0cm of c,red, dot] (c){};
	\vertex at (-2,-2) (f) {\(p_1\)};
	\vertex at (2, -2) (g) {\(p_2\)};
	\diagram*{
		(a2) --[double,thick] (a1),
		(a2) --[double] (a3),
		(b) -- [plain] (a3),
		(a3) -- [plain] (c),
		(a2) -- [boson] (b),
		(a2)-- [boson] (c) ,
		(f) -- [plain,thick] (b),
		(g) -- [plain,thick] (c),
	};
	\vertex [left=0.4em of a2] {\(3\)};
	\vertex [left=0.6em of b] {\(2\)};
	\vertex [right=0.6em of c] {\(1\)};
	\vertex [right=0.6em of a3] {\(4\)};
	\vertex[below left=0.0cm of d, red,dot] (d){};
	\vertex[below left=0.0cm of b, red,dot] (b){};
	\vertex[below left=0.0cm of c, red,dot] (c){};
	\node[below=1cm of a1, red, dot] (a2);
	\vertex[below left=0.0cm of a3, mydot] (a3){};
\end{feynman}
\end{tikzpicture}
\caption{}
\end{subfigure}%
\begin{subfigure}{.3\textwidth}
\centering
\begin{tikzpicture}[
scale=1,
mydot/.style={
	circle,
	fill=white,
	draw,
	outer sep=0pt,
	inner sep=1.5pt
}
]
\begin{feynman}
	\vertex  at (0,2) (a1){\(p_1+p_2\)};
	\node[below=1cm of a1, red, dot] (a2);
	\vertex  at (0,0) (a3); \vertex[below left=0.0cm of a3, empty dot] (a3){};
	\vertex at (0,1) (a2) ;
	\vertex at (-1, -1) (b); \vertex[below left=0.0cm of b, red,dot] (b){};
	\vertex at (1, -1) (c);  \vertex[below left=0.0cm of c,red, dot] (c){};
	\vertex at (0, -1) (d); \vertex[below left=0.0cm of d, red,dot] (d){};
	\vertex at (-2,-2) (f) {\(p_1\)};
	\vertex at (2, -2) (g) {\(p_2\)};
	\diagram*{
		(a2) --[double,thick] (a1),
		(a2) --[double] (a3),
		(b) -- [plain] (a3),
		(a3) -- [plain] (c),
		(b) -- [boson] (c),
		(a2)-- [boson, quarter right, looseness=1.5] (d) ,
		(f) -- [plain,thick] (b),
		(g) -- [plain,thick] (c),
	};
	\vertex [left=0.4em of a2] {\(3\)};
	\vertex [left=0.6em of b] {\(2\)};
	\vertex [right=0.6em of c] {\(1\)};
	\vertex [right=0.6em of a3] {\(4\)};
	\vertex[below left=0.0cm of d, red,dot] (d){};
	\vertex[below left=0.0cm of b, red,dot] (b){};
	\vertex[below left=0.0cm of c, red,dot] (c){};
	\node[below=1cm of a1, red, dot] (a2);
	\vertex[below left=0.0cm of a3, mydot] (a3){};
\end{feynman}
\end{tikzpicture}
\caption{}
\end{subfigure}%
	\caption{Feynman two-loop finite diagrams that can contribute to the chiral superpotential. Straight lines are chiral superpropagators, curved line is a vector superpropagator. Red dots denote chiral-vector vertex $2g$, white dots denote chiral vertices $\lambda$, and black dots denote anti-chiral vertices $\bar{\lambda}$.}
\label{fig:finite}
\end{figure}

In this subsection we compute finite Feynman supergraphs in detail, see Fig.\ref{fig:finite}.  These superdiagrams are finite, since they do not contain any singular subgraphs.

The first two-loop finite diagram can be easily calculated and has the following form Fig.\ref{fig:finite} a):
\beq
\begin{gathered}
\textbf{W}^{(2),A}_{fin}=\lim_{p_1,p_2\rightarrow0}g^4 (C_A-2C_F)^2\int \prod_{l=1}^5 d^8 z_l~ \lambda \Psi_1(z_1)\Phi(z_2)\Psi_2(z_3) \left\{ \frac{1}{\square_1}\delta_{2,1}  \right. \\ \left. \frac{D^2_1 \bar{D}^2_4}{16\square_4}\delta_{1,4} \frac{D^2_4 \bar D^2_3}{16\square_4} \delta_{4,3}  \frac{D^2_5}{4\square_5} \delta_{3,5}   \frac{\bar{D}^2_5 D^2_2}{16\square_5} \delta_{5,2} \frac{1}{\square_4}\delta_{4,5}  \right\}.
\end{gathered}\label{G2a}
\eeq

\begin{figure}
    \centering
        \begin{minipage}[h]{0.25\linewidth}
	\vspace{0pt}
        \includegraphics[width=\textwidth]{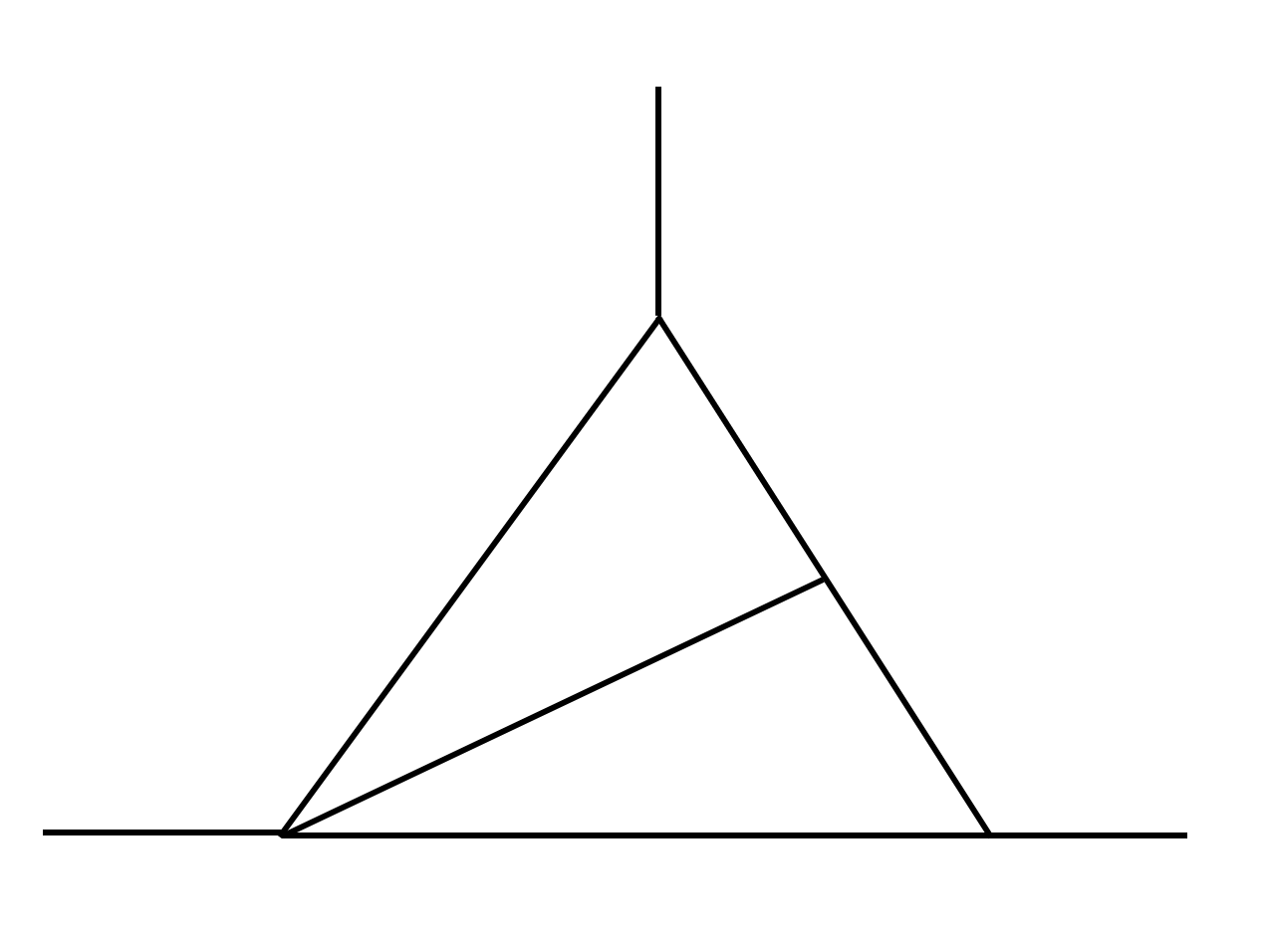}
        \label{fig:master_2}
         \end{minipage}
           \hspace{40pt}
        \begin{minipage}[h]{0.25\linewidth}
	\vspace{15pt}
        \includegraphics[width=\textwidth]{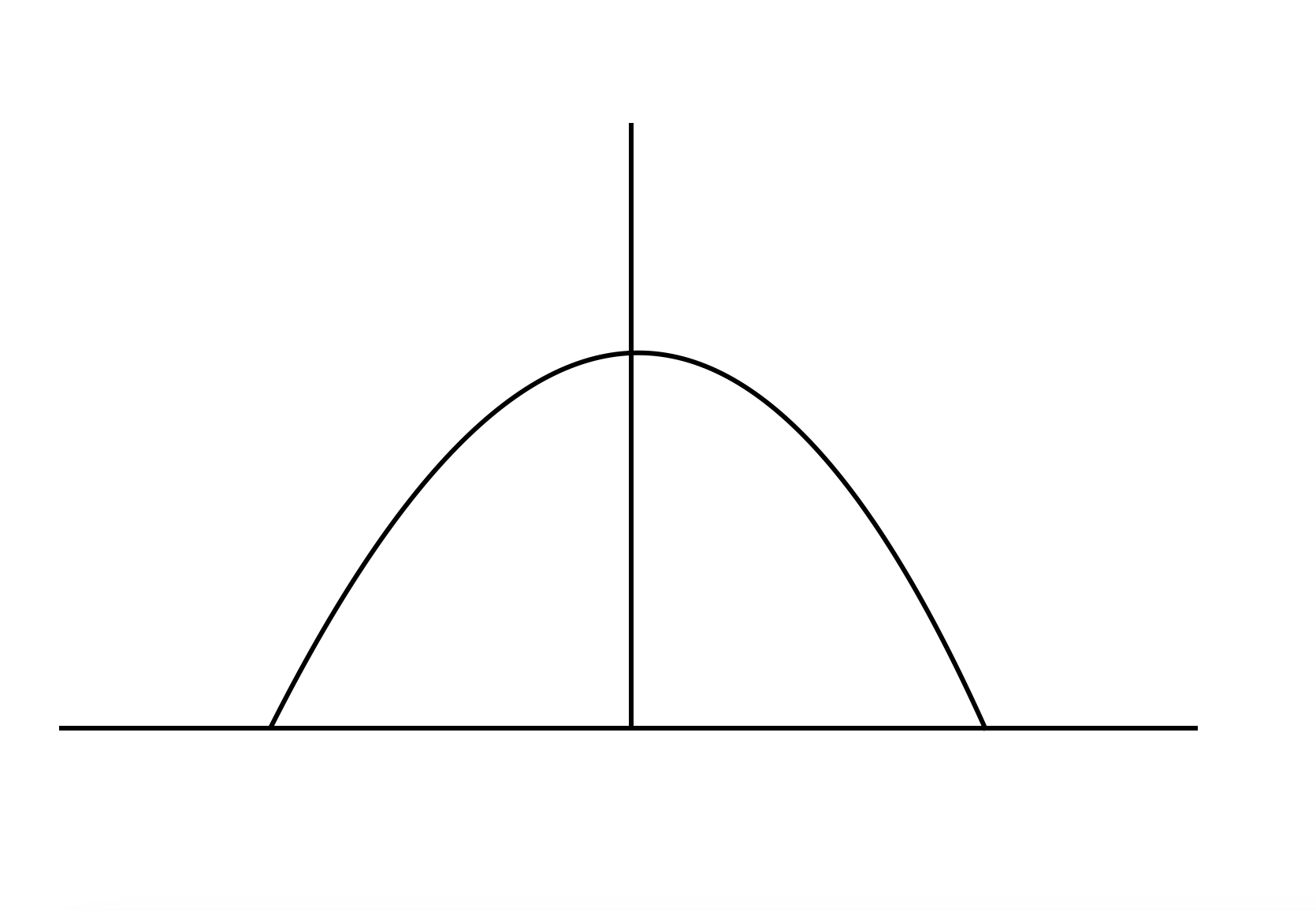}
        \label{fig:master_3}
         \end{minipage}
        \caption{Topologies of the scalar master integrals to which diagrams with finite contributions are reduced.}
\label{fig:master}
\end{figure}

After the $D$-algebra evaluation in the momentum representation one can find the integral \cite{Buchbinder:2025dgu}
\beq
\begin{gathered}
J^{(2)}_a=\lim_{p_1,p_2\rightarrow0}\int \frac{d^4 q_1}{(4\pi)^2}\frac{d^4q_2}{(4\pi)^2} \frac{q_1^2 (p_1+p_2)^2}{q_1^2 (q_1-p_1)^2 (q_2-p_2)^2 q_2^2 (q_2-p_2)^2  (q_1-p_2)^2}=6\zeta(3);
\end{gathered}
\eeq
this expression can be calculated using integration by parts \cite{Buchbinder:2025dgu} to ``walnut''-topology (see Fig.\ref{fig:master} a)) and the whole contribution can be represented as:
\beq
\textbf{W}^{(2),A}_{fin}=6g^4(C_A-2C_F)^2 \zeta(3) \times W_{tree}.
\eeq

The second diagram (Fig.\ref{fig:finite} b)) has a similar form but with non-planar internal vector lines
\beq
\begin{gathered}
\textbf{W}^{(2),B}_{fin}=\lim_{p_1,p_2\rightarrow0}4g^4(C_A-C_F)(C_A-2C_F)\int \prod_{l=1}^5 d^8 z_l  \lambda \Psi_1(z_1)\Phi(z_2)\Psi_2(z_3) \left\{ \frac{1}{\square_4}\delta_{2,4} \frac{D^2_1 \bar{D}^2_4}{16\square_4} \delta_{1,4}  \right. \\ \left. \frac{\bar{D}^2_4 D^2_3}{16\square_4} \delta_{4,3} \frac{D^2_5}{4\square_5} \delta_{3,5} \frac{\bar{D}^2_5 D^2_2}{16\square_5} \delta_{5,2}\frac{1}{\square_5}\delta_{1,5} \right\},
\end{gathered}\label{G2b}
\eeq
After the $D$-algebra evaluation in the momentum representation one can find the integral with the topology of Fig.\ref{fig:master} b) 
\beq
\begin{gathered}
J^{(2)}_b=\lim_{p_1,p_2\rightarrow0}\int \frac{d^4 q_1}{(4\pi)^2}\frac{d^4q_2}{(4\pi)^2} \frac{-q_1^2p_1^2-q_2^2p_2^2+ 2(q_1q_2)(p_1p_2)}{q_1^2 (q_1-p_1)^2 (q_1+q_2-p_1)^2 q_2^2 (q_2+p_2)^2  (q_1+q_2+p_1)^2 },
\end{gathered}
\eeq
so it can be integrated as
\beq
\textbf{W}^{(2),B}_{fin}=-24g^4(C_A-C_F)(C_A-2C_F) \zeta(3) \times W_{tree}.
\eeq

The third diagram (Fig.\ref{fig:finite} c)) can be written in the superspace notation as
\beq
\begin{gathered}
\textbf{W}^{(2),C}_{fin}=\lim_{p_1,p_2\rightarrow0}2g^4(C_A-C_F)(C_A-2C_F)\int \prod_{l=1}^5 d^8 z_l  ~\lambda  \Psi_1(z_1)\Phi(z_2)\Psi_2(z_3) \nonumber\\ \left\{ \frac{1}{\square_1}\delta_{2,1} \frac{D^2_1 \bar{D}^2_4}{16\square_4} \delta_{1,4}   \frac{\bar{D}^2_4 D^2_3}{16\square_4} \delta_{4,3} \frac{D^2_2}{4\square_3} \delta_{3,2}\frac{1}{\square_4}\delta_{2,4} \right\},
\end{gathered}\label{G2c}
\eeq
which reduces to
\beq
\begin{gathered}
J^{(2)}_c=\lim_{p_1,p_2\rightarrow0}\int \frac{d^4 q_1}{(4\pi)^2}\frac{d^4q_2}{(4\pi)^2} \frac{q_1^2(p_1+p_2)^2}{q_1^2 (q_1+p_1)^2 (q_2+p_1)^2 (q_2-q_1)^2 (q_2-p_2)^2  },
\end{gathered}
\eeq
and after integration turns to the topology of Fig.\ref{fig:master} b) , so it can be integrated in the following form
\beq
\textbf{W}^{(2),C}_{fin}=12g^4(C_A-C_F)(C_A-2C_F)\zeta(3) \times W_{tree}.
\eeq

The diagram from Fig.\ref{fig:finite} d) has superspace integral form
\beq
\begin{gathered}
\textbf{W}^{(2),D}_{fin}=\lim_{p_1,p_2\rightarrow0}2g^4 \int \prod_{l=1}^5 d^8 z_l ~ \lambda \Psi_1(z_1)\Phi(z_2)\Psi_2(z_3) \left\{ \frac{1}{\square_5}\delta_{5,1} \frac{D^2_1 \bar{D}^2_3}{16\square_1} \delta_{1,3}  \right. \\ \left.  \frac{D^2_4}{4\square_3}\delta_{3,4} \frac{\bar{D}^2_4 D^2_2}{16\square_4} \delta_{4,2}\frac{1}{\square_4}\delta_{2,5} \frac{1}{\square_4}\delta_{5,4}\right\},
\end{gathered}\label{G2d}
\eeq
 so we obtain
\beq
\begin{gathered}
J^{(2)}_d=\lim_{p_1,p_2\rightarrow0}\int \frac{d^4 q_1}{(4\pi)^2}\frac{d^4q_2}{(4\pi)^2} \frac{(p_1+p_2)^2}{q_1^2 (q_1+p_1)^2 (q_1-p_2)^2 (q_2-p_2)^2 q_2^2 (q_1-q_2)^2  },
\end{gathered}
\eeq
and the final result is
\beq
\textbf{W}^{(2),D}_{fin}=2g^4 \Upsilon^{(2)} \times W_{tree}.
\eeq
Here the scalar master-integral is a double triangle integral, see Fig.\ref{fig:ladderUD}, that can be represented in the following way:
\beq
\Upsilon^{(2)}=\int_0^1 d\tau \frac{2 \log^3(\tau)}{\tau^2-\tau+1},
\eeq
where $\Upsilon^{(2)}$ is the two-loop Usyukina-Davydychev triangle function~\eqref{eq:master1}.

The integral of Fig.\ref{fig:finite} e) has the following form:
\beq
\begin{gathered}
\textbf{W}^{(2),E}=\lim_{p_1,p_2\rightarrow0}2g^4(C_A-2C_F)^2\int \prod_{l=1}^5 d^8 z_l ~\lambda \Psi_1(z_1)\Phi(z_2)\Psi_2(z_3) \left\{ \frac{1}{\square_4}\delta_{1,4} \frac{D^2_1 \bar{D}^2_5}{16\square_5} \delta_{1,5}  \right. \\ \left. \frac{D^2_5 \bar{D}^2_4}{16\square_5} \delta_{5,4} \frac{D^2_4 \bar{D}^2_3}{16\square_4} \delta_{4,3} \frac{D^2_2}{4\square_3} \delta_{3,2}\frac{1}{\square_5}\delta_{2,5} \right\}.
\end{gathered}\label{G2e}
\eeq

After the $D$-algbra evaluation, we have the master integral with the topology of Fig.\ref{fig:master} b) 
\beq
\begin{gathered}
J^{(2)}_e=\lim_{p_1,p_2\rightarrow0}\int \frac{d^4 q_1}{(4\pi)^2}\frac{d^4q_2}{(4\pi)^2} \frac{-q_1^2p_1^2+(q_1-q_2)^2p_2^2-2(q_1q_2)(p_1p_2)}{q_1^2 (q_1-p_1)^2 (q_2-q_1+p_1)^2 (q_2-p_2)^2 q_2^2 (q_2+p_1)^2  },
\end{gathered}
\eeq
which is again equal to $6\zeta(3)$, so the final result for the integral can be expressed as
\beq
\textbf{W}^{(2),E}_{fin}(\Phi)=-12g^4(C_A-2C_F)^2\zeta(3) \times W_{tree}.
\eeq

The diagram in Fig.\ref{fig:finite} f) can be written as follows:
\beq
\begin{gathered}
\textbf{W}^{(2),F}_{fin}(\Phi)\sim\lim_{p_1,p_2\rightarrow0} 4g^4\int \prod_{l=1}^5 d^8 z_l  \lambda \Psi_1(z_1)\Phi(z_2)\Psi_2(z_3) \left\{ \frac{1}{\square_4}\delta_{2,4} \frac{D^2_1 \bar{D}^2_4}{16\square_4} \delta_{1,4}  \right. \\ \left. \frac{\bar{D}^2_4 D^2_3}{16\square_4} \delta_{4,3} \frac{D^2_5}{4\square_5} \delta_{3,5} \frac{\bar{D}^2_5 D^2_2}{16\square_5} \delta_{5,2}\frac{1}{\square_5}\delta_{1,5} \right\},
\end{gathered}\label{G2f}
\eeq
and after the $D$-algebra calculation one can reduce it to
\beq
\begin{gathered}
J^{(2)}_f=0.
\end{gathered}
\eeq
The reason for this diagram after $D$-algebra transformations produces the $\bar{D}^2\Phi$ term which is identically zero.

For Fig.\ref{fig:finite} g) we have
\beq
\begin{gathered}
\textbf{W}^{(2),G}_{fin} \sim \lim_{p_1,p_2\rightarrow0}4g^4\int \prod_{l=1}^5 d^8 z_l  \lambda \Psi_1(z_1)\Phi(z_2)\Psi_2(z_3) \left\{ \frac{D^2_1 \bar{D}^2_4}{16\square_4}\delta_{1,4} \frac{\bar{D}^2_4 D^2_2}{16\square_4} \delta_{4,2}  \right. \\ \left. \frac{1}{\square_3} \delta_{2,3} \frac{1}{\square_1} \delta_{3,1} \frac{D^2_3}{4\square_4} \delta_{3,4} \right\}=0,
\end{gathered}\label{G2g}
\eeq
and the last diagram
\beq
\begin{gathered}
\textbf{W}^{(2),H}\sim\lim_{p_1,p_2\rightarrow0}2g^4\int \prod_{l=1}^5 d^8 z_l ~\lambda  \Psi_1(z_1)\Phi(z_2)\Psi_2(z_3) \left\{ \frac{1}{\square_5} \delta_{3,5}\frac{D^2_1 \bar{D}^2_4}{16\square_4}\delta_{1,4} \frac{\bar{D}^2_4 D^2_2}{16\square_4} \delta_{4,2}  \right. \\ \left. \frac{1}{\square_3} \delta_{2,5} \frac{1}{\square_1} \delta_{5,3} \frac{D^2_3}{4\square_4} \delta_{3,4} \right\}=0.
\end{gathered}\label{G2h}
\eeq
Both master-integrals~\eqref{G2g}-\eqref{G2h} are equal to zero here due to the adopted Fermi-Feynman gauge.

So finally restoring all factors of the $SU(N)$ gauge group, we have the following result for the two-loop finite chiral effective superpotential (see Fig.\ref{Fig:twoloopchiral} and Fig.\ref{fig:finite})
\begin{equation}
	\textbf{W}^{(2)}_{fin} =\left[2g^4\Upsilon^{(2)} + 6(C_A-2C_F)\left(2(C_A-C_F)|\lambda|^4 -(3 C_A-4C_F)g^4\right) \zeta(3)\right] \times W_{tree}
	\label{finWN1}
\end{equation}
so for the $\mathcal{N}=2$ SYM result is simplified to
\begin{equation}
	\textbf{W}^{(2)}_{fin,\mathcal{N}=2} =2g^4\left(\Upsilon^{(2)} -3\frac{1}{N^2} \zeta(3)\right) \times W_{tree}
	\label{finWN2}
\end{equation}
Unexpectedly that only one diagram has leading contribution by $N$, it corresponds to Fig.\ref{fig:finite} d). Note that these gauge-containing and pure 'chiral' terms are reduced only due to the specifics of the interaction between chiral and gauge superfields. As for \eqref{finWN1}, we see that due to the coupling coefficient $\lambda$ in the large $N$ limit, there are more terms contributing to the chiral superpotential. 

It can be seen that all the above two-loop diagrams except for the triangle-type diagrams, Fig.\ref{fig:finite} a), reduce to the well-known walnut-type graph topology (see Fig.\ref{fig:master} b). This peculiarity can be used to study various limits of the theory, for example, in the  large $N$ limit of the $SU(N)$ gauge group.
Note that similar contributions can be derived for other super Yang-Mills models, for instance, $\beta$-deformed $\mathcal{N}=1$ SYM, etc.\cite{tobepublished}.

\subsection{Final result}\label{sec:fin_res}

Thus, one can finally write the sum of two contributions to the model from \eqref{N1Wdiv} and \eqref{finWN1}:
\begin{equation}
	\begin{gathered}
		\textbf{W}^{(2)}=\textbf{W}^{(2)}_{div}+\textbf{W}^{(2)}_{fin}=\left[
		\left(g^4\left\{2N_fT_F-N\right\}+g^2(|\lambda^2|-g^2)\right)J^{(1)}_{1,1}\Upsilon^{(1)}+\right. \\ \left.~  \frac{1}{2}g^4\Upsilon^{(2)}+ \left(-3(C_A-C_F))|\lambda|^4 +\frac{3}{2}(3C_A-4C_F)g^4\right)\zeta(3)\right]\times 4(2C_F-C_A)W_{tree}
	\end{gathered}
\end{equation}
This expression contains the UV singular contribution, but as the considered action is renormalizable, the divergences can be absorbed into the redefinitions of the coupling constants and fields.
It is known that bare and renormalized quantities in supersymmetric theories are related as (see e.g. \cite{Superspace:1983nr,WestBook})
\begin{equation}
\begin{gathered}
\Phi = (z_\phi^{1/2})\Phi_R,~ \bar{\Phi} = (z^{1/2}_\phi)\bar{\Phi}_R\\
\Psi_I= (z^{1/2}_\psi)\Psi_{I,R},~ \bar{\Phi} = (z^{1/2}_\psi)\bar{\Psi}_{I,R} \\
V= z_V^{1/2}V_R,~ g = z_g g_R,~\lambda=z_\lambda^{3/2} \lambda  \label{renormalized}
\end{gathered}
\end{equation}
Despite the fact that we excluded the contribution with the external gauge superfields $V$ and concentrated only on the chiral superfield part, the gauge superfield one-loop divergent contribution was taken into account in our calculations. Thus, we need to renormalize the gauge coupling constant $g$. The $\mathcal{N}=1$ non-renormalization theorem requires that the chiral superfield part of the Lagrangian is not renormalized, so we obtain
\begin{equation}
    z_\phi^{1/2}= z^{-1}_\psi
\end{equation}
Replacing now the bare fields and the bare coupling with the renormalized ones, according to Eq.~\eqref{renormalized}, we get rid of the UV singularity and obtain the renormalized expression in terms of renormalized quantities
\begin{align}
	\textbf{W}^{(2)}_R=&
	\left( \frac{1}{2}g_R^4\Upsilon^{(2)}+\left(-3(C_A-C_F)|\lambda|_R^4 +\frac{3}{2}(3C_A-4C_F)g_R^4\right) \zeta(3)+2u\right)\nonumber\\ &\times 4(2C_F-C_A)W_{tree,R}.
	\label{Wtwoloop2R}
\end{align}
where $W_{tree,R}=\text{tr}\int d^6z~\lambda_R \Psi_{1,R}\Phi_R \Psi_{2,R}$ and $u=\left(g_R^4\left\{2N_fT_F-C_A\right\}+g_R^2(|\lambda_R^2|-g_R^2)\right)\Upsilon^{(1)}$ corresponding to the constant part left after the substraction of singularities. This is the final result of our two-loop level calculation.

\section{Large $N$ limit in chiral effective superpotential for finite $\mathcal{N}=2$ SYM}\label{sec:largeN}

Let us assume that we have the $SU(N)$ gauge group and single coupling in the theory requiring $\lambda=g$ which restores extended $\mathcal{N}=2$ supersymmetry.
It is clear that if the $\mathcal{N}=2$  SYM theory is finite at the one-loop level, the quantum corrections to the effective action will be determined by finite chiral corrections. Thus, we concentrate on a class of finite SYM theories and further consider only loop chiral effective potentials for finite theories.

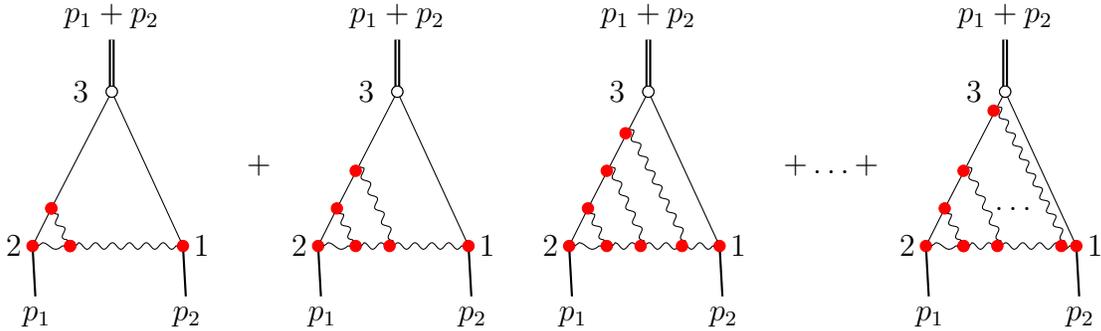
\begin{figure}[htb]
\centering
        \begin{minipage}[h]{0.2\linewidth}
	\vspace{0pt}
        \begin{subfigure}{.25\textwidth}
        	\centering
        	\begin{tikzpicture}[
        		scale=1,
        		mydot/.style={
        			circle,
        			fill=white,
        			draw,
        			outer sep=0pt,
        			inner sep=1.5pt
        		}
        		]
        		\begin{feynman}
        			\vertex  at (0,2) (a1){\(p_1+p_2\)};
        			\node[below=1cm of a1, mydot] (a2);
        			\vertex at (0,1) (a2) ;
        			\vertex at (-0.5,0) (q); %\vertex[below left=0.0cm of b, red,dot] (b){};
        			\vertex at (0, -1) (w);  %\vertex[below left=0.0cm of c,red, dot] (c){};
        			\vertex at (-0.75, -0.5) (b); \vertex[below left=0.0cm of b, red,dot] (b){};
        			\vertex at (-0.5, -1) (c);  \vertex[below left=0.0cm of c,red, dot] (c){}; 
        			\vertex at (-1, -1) (d); \vertex[below left=0.0cm of d, red, dot] (d){};
        			\vertex at (1, -1) (e); \vertex[below left=0.0cm of e, red,dot] (e){};
        			\vertex at (-1,-2) (f) {\(p_1\)};
        			\vertex at (1, -2) (g) {\(p_2\)};
        			\diagram*{
        				(a2) --[double,thick] (a1),
        				(b) -- [plain] (a2),
        				(a2) -- [plain] (e),
        				%(a2) -- [plain] (c),
        				(c) -- [boson] (b),
        				%(cc) -- [boson] (bb),
        				(e) -- [boson] (c),
        				(c)-- [boson] (d) ,
        				(d) -- [plain] (b),
        				%(e) -- [boson, momentum=\(k\)] (d) ,
        				(f) -- [plain,thick] (d),
        				(g) -- [plain,thick] (e),
        			};
        			\vertex [left=0.4em of a2] {\(3\)};
        			\vertex [left=0.6em of d] {\(2\)};
        			\vertex [right=0.6em of e] {\(1\)};
        			\node[below=1cm of a1, mydot] (a2);
        			 \vertex[below left=0.0cm of b, red,dot] (b){};
        			 \vertex[below left=0.0cm of c,red, dot] (c){};
        		\end{feynman}
        	\end{tikzpicture}
        	 \end{subfigure}
         \end{minipage}
         $+$
        \begin{minipage}[h]{0.2\linewidth}
	\vspace{0pt}
       \begin{subfigure}{.25\textwidth}
       	\centering
       	\begin{tikzpicture}[
       		scale=1,
       		mydot/.style={
       			circle,
       			fill=white,
       			draw,
       			outer sep=0pt,
       			inner sep=1.5pt
       		}
       		]
       		\begin{feynman}
       			\vertex  at (0,2) (a1){\(p_1+p_2\)};
       			\node[below=1cm of a1,mydot] (a2);
       			\vertex at (0,1) (a2) ;
       			\vertex at (-0.5,0) (q); 
       			\vertex at (0, -1) (w);  
       			\vertex at (-0.75, -0.5) (b); \vertex[below left=0.0cm of b, red,dot] (b){};
       			\vertex at (-0.25, 0.5) (bb);
       			\vertex at (-0.5, 0) (bbb); 
       			\vertex at (-0.5, -1) (c);  \vertex[below left=0.0cm of c,red, dot] (c){};
       			\vertex at (0.5, -1) (cc); 
       			\vertex at (-0.05, -1) (ccc);  
       			\vertex at (-1, -1) (d); \vertex[below left=0.0cm of d, red, dot] (d){};
       			\vertex at (1, -1) (e); \vertex[below left=0.0cm of e, red,dot] (e){};
       			\vertex at (-1,-2) (f) {\(p_1\)};
       			\vertex at (1, -2) (g) {\(p_2\)};
       			\diagram*{
       				(a2) --[double,thick] (a1),
       				(b) -- [plain] (a2),
       				(a2) -- [plain] (e),
       				%(a2) -- [plain] (c),
       				(c) -- [boson] (b),
       				%	(cc) -- [boson] (bb),
       				(ccc) -- [boson] (bbb),
       				(e) -- [boson] (c),
       				(c)-- [boson] (d) ,
       				(d) -- [plain] (b),
       				%(e) -- [boson, momentum=\(k\)] (d) ,
       				(f) -- [plain,thick] (d),
       				(g) -- [plain,thick] (e),
       			};
       			\vertex [left=0.4em of a2] {\(3\)};
       			\vertex [left=0.6em of d] {\(2\)};
       			\vertex [right=0.6em of e] {\(1\)};
       			\node[below=1cm of a1, mydot] (a2);
       			\vertex[below left=0.0cm of ccc,red, dot] (ccc){};
       			\vertex[below left=0.0cm of bbb, red,dot] (bbb){};
       		\end{feynman}
       	\end{tikzpicture}
       	%	\caption{}
       \end{subfigure}%
         \end{minipage}
          \begin{minipage}[h]{0.2\linewidth}
	\vspace{0pt}
         \begin{subfigure}{.25\textwidth}
	\centering
	\begin{tikzpicture}[
		scale=1,
		mydot/.style={
			circle,
			fill=white,
			draw,
			outer sep=0pt,
			inner sep=1.5pt
		}
		]
		\begin{feynman}
			\vertex  at (0,2) (a1){\(p_1+p_2\)};
			\node[below=1cm of a1, mydot] (a2);
			\vertex at (0,1) (a2) ;
			\vertex at (-0.5,0) (q); %\vertex[below left=0.0cm of b, red,dot] (b){};
			\vertex at (0, -1) (w);  %\vertex[below left=0.0cm of c,red, dot] (c){};
			\vertex at (-0.75, -0.5) (b); \vertex[below left=0.0cm of b, red,dot] (b){};
			\vertex at (-0.25, 0.5) (bb);
			\vertex at (-0.5, -1) (c);  \vertex[below left=0.0cm of c,red, dot] (c){};
			\vertex at (0.5, -1) (cc);  
			\vertex at (-1, -1) (d); \vertex[below left=0.0cm of d, red, dot] (d){};
			\vertex at (1, -1) (e); \vertex[below left=0.0cm of e, red,dot] (e){};
	                \vertex at (-0.5, 0) (bbb); 
			\vertex at (-0.05, -1) (ccc); 
			\vertex at (-1,-2) (f) {\(p_1\)};
			\vertex at (1, -2) (g) {\(p_2\)};
			\diagram*{
				(a2) --[double,thick] (a1),
				(b) -- [plain] (a2),
				(a2) -- [plain] (e),
				%(a2) -- [plain] (c),
				(c) -- [boson] (b),
				(cc) -- [boson] (bb),
				(ccc) -- [boson] (bbb),
				(e) -- [boson] (c),
				(c)-- [boson] (d) ,
				(d) -- [plain] (b),
				%(e) -- [boson, momentum=\(k\)] (d) ,
				(f) -- [plain,thick] (d),
				(g) -- [plain,thick] (e),
			};
            \vertex [left=0.4em of a2] {\(3\)};
        \vertex [left=0.6em of d] {\(2\)};
        \vertex [right=0.6em of e] {\(1\)};
         \vertex[below left=0.0cm of bbb, red,dot] (bbb){};
          \vertex[below left=0.0cm of ccc,red, dot] (ccc){};
          \vertex[below left=0.0cm of cc,red, dot] (cc){};
           \vertex[below left=0.0cm of bb, red,dot] (bb){};
        		\node[below=1cm of a1, mydot] (a2);
		\end{feynman}
	\end{tikzpicture}
\end{subfigure}%
          \end{minipage}
         $+ \ldots +$
        \begin{minipage}[h]{0.2\linewidth}
	\vspace{0pt}
             \begin{subfigure}{.25\textwidth}
       	\centering
       	\begin{tikzpicture}[
       		scale=1,
       		mydot/.style={
       			circle,
       			fill=white,
       			draw,
       			outer sep=0pt,
       			inner sep=1.5pt
       		}
       		]
       		\begin{feynman}
       			\vertex  at (0,2) (a1){\(p_1+p_2\)};
       			\node[below=1cm of a1, mydot] (a2);
       			\vertex at (0,1) (a2) ;
       			\vertex at (-0.5,0) (q); %\vertex[below left=0.0cm of b, red,dot] (b){};
       			\vertex at (0, -1) (w);  \vertex[below left=0.0cm of c,red, dot] (c){};
       			\vertex at (-0.75, -0.5) (b); \vertex[below left=0.0cm of b, red,dot] (b){};
       			\vertex at (-0.1, 0.8) (bb);
       			\vertex at (-0.5, 0) (bbb); 
       			\vertex at (-0.5, -1) (c);  \vertex[below left=0.0cm of c,red, dot] (c){};
       			\vertex at (0.8, -1) (cc);  
       			\vertex at (-0.05, -1) (ccc);  
       			\vertex at (-1, -1) (d); \vertex[below left=0.0cm of d, red, dot] (d){};
       			\vertex at (1, -1) (e); \vertex[below left=0.0cm of e, red,dot] (e){};
       			\vertex at (-1,-2) (f) {\(p_1\)};
       			\vertex at (1, -2) (g) {\(p_2\)};
       			\diagram*{
       				(a2) --[double,thick] (a1),
       				(b) -- [plain] (a2),
       				(a2) -- [plain] (e),
       				%(a2) -- [plain] (c),
       				(c) -- [boson] (b),
       				(cc) -- [boson] (bb),
       				(ccc) -- [boson] (bbb),
       				(e) -- [boson] (c),
       				(c)-- [boson] (d) ,
       				(d) -- [plain] (b),
       				%(e) -- [boson, momentum=\(k\)] (d) ,
       				(f) -- [plain,thick] (d),
       				(g) -- [plain,thick] (e),
       			};
       			\vertex [left=0.4em of a2] {\(3\)};
       			\vertex [right=2.2em of b] {\(\ldots\)};
       			\vertex [left=0.6em of d] {\(2\)};
       			\vertex [right=0.6em of e] {\(1\)};
       			\node[below=1cm of a1, mydot] (a2);
       			\vertex[below left=0.0cm of bb, red,dot] (bb){};
       			\vertex[below left=0.0cm of bbb, red,dot] (bbb){};
       			\vertex[below left=0.0cm of cc,red, dot] (cc){};
       			\vertex[below left=0.0cm of ccc,red, dot] (ccc){};
       		\end{feynman}
       	\end{tikzpicture}
       \end{subfigure}%
         \end{minipage}
\caption{Sum of a series of triangle-type diagrams. }
\label{fig:ladder0}
\end{figure}

In the previous sections we saw that even in the considered $\mathcal{N}=1$ SYM theory some parts of two-loop corrections giving the contribution to the chiral superpotential are canceled. Also, it could be noticed that the remaining parts in one-loop and second-loop order have a similar structure;  these parts are also leading in the number of colors $N$. We will refer to these specific contributions as a series of triangle ladder diagrams. So in this section we consider a series of ladder triangle diagrams (see, e.g. Fig.\ref{fig:ladder0})  in the limit of the large number of colors $N$ ($N\to \infty$) in the $SU(N)$ gauge $\mathcal{N}=2$ super-Yang-Mills theory.

 \begin{figure}
    \centering
\begin{minipage}[h]{0.2\linewidth}
	\vspace{0pt}
      \begin{subfigure}{.25\textwidth}
      	\centering
      	\begin{tikzpicture}[
		scale=1,
		mydot/.style={
			circle,
			fill=white,
			draw,
			outer sep=0pt,
			inner sep=1.5pt
		}
		]
		\begin{feynman}
			\vertex  at (0,2) (a1){\(p_1+p_2\)};
			\node[below=1cm of a1, mydot] (a2);
			\vertex at (0,1) (a2) ;
			\vertex at (-0.5,0) (q); %\vertex[below left=0.0cm of b, red,dot] (b){};
			\vertex at (0, -1) (w);  %\vertex[below left=0.0cm of c,red, dot] (c){};
			\vertex at (-0.75, -0.5) (b); \vertex[below left=0.0cm of b, red,dot] (b){};
			\vertex at (-0.25, 0.5) (bb); 
			\vertex at (-0.5, -1) (c);  \vertex[below left=0.0cm of c,red, dot] (c){};
			\vertex at (0.5, -1) (cc);  
			\vertex at (-1, -1) (d); \vertex[below left=0.0cm of d, red, dot] (d){};
			\vertex at (1, -1) (e); \vertex[below left=0.0cm of e, red,dot] (e){};
			\vertex at (-1,-2) (f) {\(p_1\)};
			\vertex at (1, -2) (g) {\(p_2\)};
			\diagram*{
				(a2) --[double,thick] (a1),
				(b) -- [plain] (a2),
				(a2) -- [plain] (e),
				%(a2) -- [plain] (c),
				(c) -- [boson] (b),
				(cc) -- [boson] (bb),
				(e) -- [boson] (c),
				(c)-- [boson] (d) ,
				(d) -- [plain] (b),
				%(e) -- [boson, momentum=\(k\)] (d) ,
				(f) -- [plain,thick] (d),
				(g) -- [plain,thick] (e),
			};
            \vertex [left=0.4em of a2] {\(3\)};
               \vertex [right=1.2em of b] {\(\ldots\)};
    %    \vertex [left=0.6em of b] {\(4\)};
      %  \vertex [below=0.6em of c] {\(5\)};
        \vertex [left=0.6em of d] {\(2\)};
        \vertex [right=0.6em of e] {\(1\)};
        		\node[below=1cm of a1, mydot] (a2);
		\vertex[below left=0.0cm of cc,red, dot] (cc){};
		\vertex[below left=0.0cm of bb, red,dot] (bb){};
		\end{feynman}
	\end{tikzpicture}      \end{subfigure}%
         \end{minipage}$\to$
        \begin{minipage}[h]{0.25\linewidth}
	\vspace{0pt}
        \includegraphics[width=\textwidth]{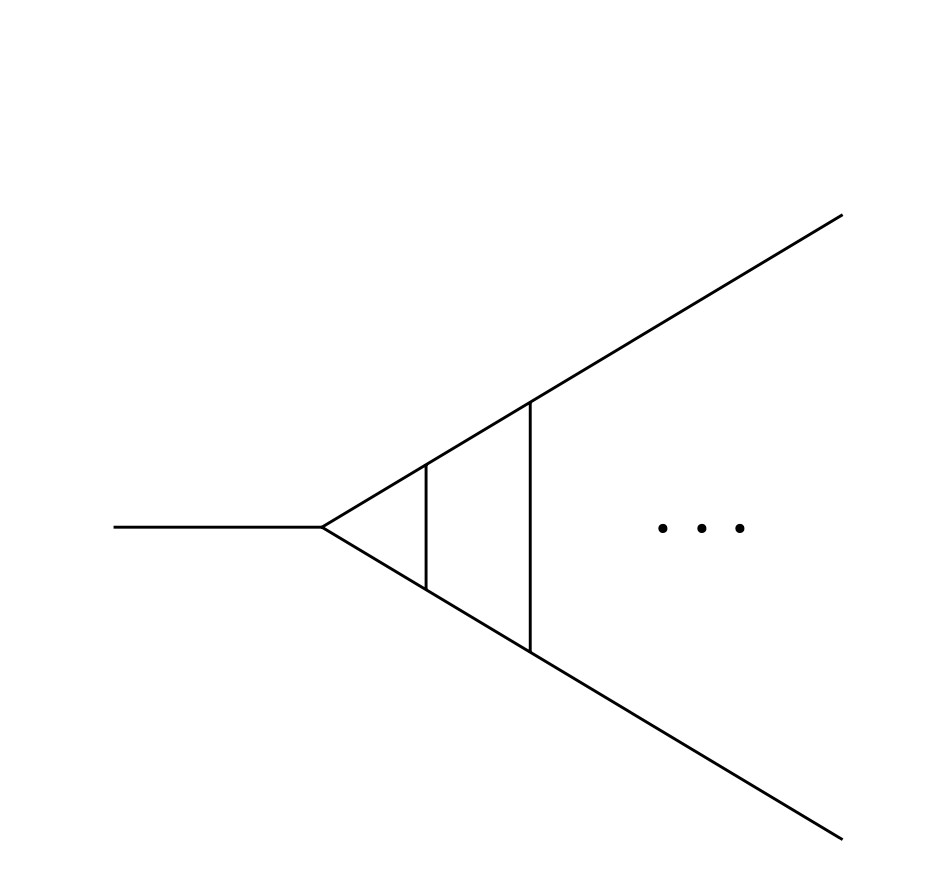}
         \end{minipage}
        \caption{A sequence of triangle type scalar master diagrams obtained from the reduction of supergraphs that can be summed up exactly in $\mathcal{N}=2$ SYM theory.}
\label{fig:ladderUD}
\end{figure}
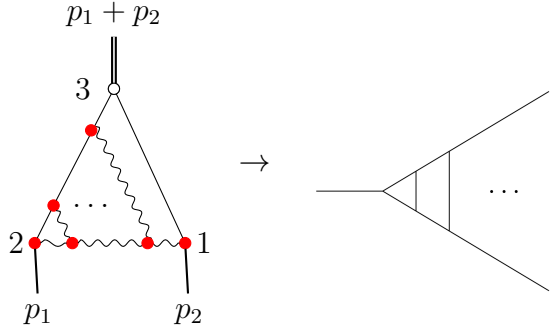

It turns out that we can sum the full set of triangle-type diagrams and reduce them to the sequence of scalar master diagrams.
We manually checked these diagrams from the first two-loops described in the previous section up to the fifth loop order utilizing \texttt{SusyMath.m}\cite{Ferrari:2007sc} and found their predictive structure. In the following we discuss these diagrams in more detail.

Let us consider the diagram depicted in Fig. \ref{fig:ladderUD} on the right panel:
\begin{equation}
\begin{gathered}
    \textbf{W}'^{(m)}=\lim_{p_1,p_2\rightarrow0}(-1)^{m+1}\frac{g^{2m}}{4} N^{m-2} \int \prod_{l=1}^{2m+1} d^8 z_l ~ \lambda \Psi_1(z_1)\Phi(z_2)\Psi_2(z_3) \times \\ \times\left\{ \frac{1}{\square_{5}}\delta_{5,1} \frac{D^2_1 \bar{D}^2_3}{16\square_1} \delta_{1,3} \frac{D^2_{4}}{4\square_3}\delta_{3,4} \frac{\bar{D}^2_4 D^2_6}{16\square_4}  \delta_{4,6}\ldots \right. \\ \left.  \dots\frac{\bar{D}_{2m-2}^2 D_{2m}^2}{\square_{2m-2}}\delta_{2m-2,2m}\frac{\bar{D}^2_{2m}D^2_2}{\square_{2m}}\delta_{2m,2}\frac{1}{\square_{2m+1}}\delta_{2m+1,2}\ldots \right. \\ \left.  \frac{1}{\square_{2m}}\delta_{2m+1,2m} \ldots \frac{1}{\square_4}\delta_{5,4}\right\}.
\end{gathered}
\end{equation}
The numerator for this diagram is also quite simple to compute: first, it is necessary to remove all free delta functions over Grassmann variables and use the relations for covariant derivatives listed in Appendix \ref{sec:AppA}.
After evaluating $D$-algebra and re-expressing the integral through the $\Upsilon$-functions, the result for this diagram in the general form can be written as:
\begin{equation}
\textbf{W}'^{(m)}=(-1)^{m-1}\frac{g^{2m}}{4} N^{m-2} \Upsilon^{(m)} \times W_{tree}.
\end{equation}
This expression generalizes the finite contributions found in the previous section.
The formal sums $\textbf{W}'^{lead}=\sum \textbf{W}'^{(m)} $ over these contributions (assuming $y \log (\tau)\leq1$ which is justified in the large $N$ limit where $y=g^2 N$) are given in compact form after taking the limit $p_{1},p_2\rightarrow0$ in the following expression:
\begin{equation}
\textbf{W}'^{lead}=\frac{y}{2N^2}\int_0^1~d \tau \frac{\log(\tau)~(1-\tau)}{(1+y \log^2(\tau))~(1+\tau^3) } \times W_{tree}=\Upsilon^{tot} \times W_{tree},
\end{equation}
and after integration one has the closed result
\begin{equation}
    \Upsilon^{tot}=\frac{1}{4N^2}\sum_{m=1}^\infty ((\pi -2 \text{Si}(x)) \sin (x)-2 \text{Ci}(x) \cos (x))U_m\left(1/2\right),
\end{equation}
where $x=\frac{m+1}{\sqrt{y}}$ and $\text{Si}$ and $\text{Ci}$ are the integral sine and cosine functions and $U_{n}(x)$ are Chebyshev polynomials.  That is, this expression exactly sums up all the leading color contributions to the chiral effective superpotential of the finite $\mathcal{N}=2$ SYM theory.

For instance, the last expression is quite convenient for studying the behaviour of the full summed expression of the chiral superpotential at large  $\hbar$. At large coupling and in the large $N$ limit, restoring $\hbar$, one can obtain 
\begin{equation}
\Upsilon^{tot}\big|_{\hbar \rightarrow \infty }\simeq \frac{1}{2 N^2}\log \left(\frac{12e^{\gamma_E }}{\sqrt{\hbar} g N} \frac{\Gamma \left(\frac{2}{3}\right) \Gamma \left(\frac{5}{6}\right)}{\Gamma \left(\frac{1}{6}\right) \Gamma \left(\frac{1}{3}\right)}\right)+\text{O}\left(\hbar^{-2}\right),
\end{equation}
where $\gamma_E$ is the Euler constant. As one can see, the classical limit of this expression does not exist and the effective superpotential depends on the coupling constant non-polynomially. We can note that this expression is similar to the leading behaviour of the continuum limit of fishnet-like graphs at `critical coupling'. It is quite possible that there are connections between the concepts of the critical coupling constant in the thermodynamic limit of  fishnet models and the constant in the considered limit for chiral effective potentials \cite{Basso:2018agi,Basso:2019xay}. It would be extremely interesting to demonstrate the applicability of the fishnet model methods to computations of various types of object in supersymmetric models, as it was done, e.g. in \cite{Kade:2024lkc,Kade:2024ucz}.

\begin{figure}[htb]
\centering
		 \begin{minipage}[h]{0.2\linewidth}
	\vspace{0pt}
      \begin{subfigure}{.3\textwidth}
      	\centering
      	\begin{tikzpicture}[
      		scale=1,
      		mydot/.style={
      			circle,
      			fill=white,
      			draw,
      			outer sep=0pt,
      			inner sep=1.5pt
      		}
      		]
      		\begin{feynman}
      			\vertex  at (0,2) (a1){\(p_1+p_2\)};
      			\node[below=1cm of a1, mydot] (a2);
      			\vertex at (0,1) (a2) ;
      			\vertex at (-0.5, 0) (b); %\vertex[below left=0.5cm of b, red,dot] (b){};
      			\vertex at (0.5, 0) (c);  %\vertex[below left=0.5cm of c,red, dot] (c){};
      			\vertex at (-0.25, 0.5) (bb); 
      			\vertex at (0.25, 0.5) (cc);  
      			\vertex at (-0.75, -0.5) (bbb); 
      			\vertex at (0.75, -0.5) (ccc);  
      			\vertex at (-1, -1) (d); \vertex[below left=0.0cm of d, red, dot] (d){};
      			\vertex at (1, -1) (e); \vertex[below left=0.0cm of e, red,dot] (e){};
      			\vertex at (-1,-2) (f) {\(p_1\)};
      			\vertex at (1, -2) (g) {\(p_2\)};
      			\diagram*{
      				(a2) --[double,thick] (a1),
      				(b) -- [plain] (a2),
      				(a2) -- [plain] (c),
      				%    (a2) -- [plain] (ccc),
      				%   (c)-- [boson] (b) ,
      				(cc)-- [boson] (bb) ,
      				(ccc)-- [boson] (bbb) ,
      				(d) -- [plain] (b),
      				(c) -- [plain] (e),
      				(d) -- [plain] (b),
      				(e) -- [boson] (d) ,
      				(f) -- [plain,thick] (d),
      				(g) -- [plain,thick] (e),
      			};
      			\vertex [left=0.4em of c] {\(\ldots\)};
      			\node[below=1cm of a1, mydot] (a2);
      			\vertex[below left=0.0cm of bb, red,dot] (bb){};
      			\vertex[below left=0.0cm of cc,red, dot] (cc){};
      			\vertex[below left=0.0cm of bbb, red,dot] (bbb){};
      			\vertex[below left=0.0cm of ccc,red, dot] (ccc){};
      		\end{feynman}
      	\end{tikzpicture}
      \end{subfigure}
       \end{minipage} $\to$
         \begin{minipage}[h]{0.2\linewidth}
	\vspace{0pt}
        \includegraphics[width=1\linewidth]{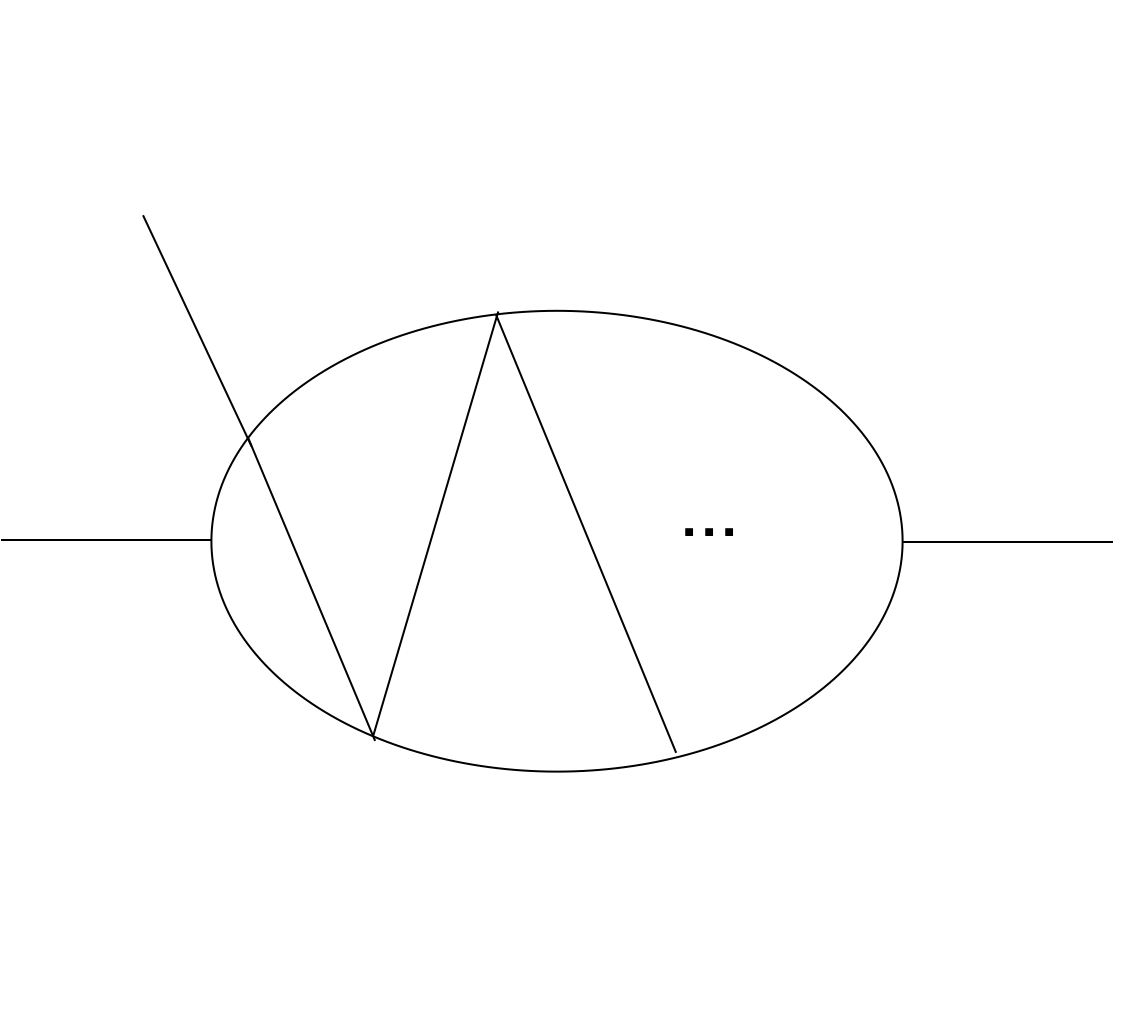}
        \end{minipage}
        		\begin{minipage}[h]{0.2\linewidth}
	\vspace{0pt}
        \begin{subfigure}{.3\textwidth}
        	\centering
        	\begin{tikzpicture}[
        		scale=1,
        		mydot/.style={
        			circle,
        			fill=white,
        			draw,
        			outer sep=0pt,
        			inner sep=1.5pt
        		}
        		]
        		\begin{feynman}
        			\vertex  at (0,2) (a1){\(p_1+p_2\)};
        			\node[below=1cm of a1, mydot] (a2);
        			\vertex at (0,1) (a2) ;
        			%	\vertex at (-1, 0) (b); \vertex[below left=0.0cm of b, dot] (b){};
        			\vertex at (0.75, -0.5) (b); \vertex[below left=0.0cm of b, red,dot] (b){};
        			\vertex at (0.25, 0.5) (bb); 
        			\vertex at (-1, -1) (d); \vertex[below left=0.0cm of d, red, dot] (d){};
        			\vertex at (1, -1) (e); \vertex[below left=0.0cm of e, red,dot] (e){};
        			\vertex at (-1,-2) (f) {\(p_1\)};
        			\vertex at (1, -2) (g) {\(p_2\)};
        			\diagram*{
        				(a2) --[double,thick] (a1),
        				(d) -- [plain] (a2),
        				(a2) -- [plain] (e),
        				%(a2) -- [plain] (c),
        				(e)-- [boson] (d) ,
        				(d) -- [boson] (b),
        				(d) -- [boson] (bb),
        				%(c) -- [plain] (e),
        				%	(d) -- [plain] (b),
        				%	(e) -- [boson, momentum=\(k\)] (d) ,
        				(f) -- [plain,thick] (d),
        				(g) -- [plain,thick] (e),
        			};
        			\vertex [left=0.4em of a2] {\(3\)};
        			%   \vertex [right=0.6em of c] {\(4\)};
        			\vertex [below=1.5em of bb] {\(\ldots\)};
        			\vertex [left=0.6em of d] {\(2\)};
        			\vertex [right=0.6em of e] {\(1\)};
        			\node[below=1cm of a1, mydot] (a2);
        			\vertex[below left=0.0cm of bb, red,dot] (bb){};
        			 \vertex[below left=0.0cm of b, red,dot] (b){};
        		\end{feynman}
        	\end{tikzpicture}
        \end{subfigure}
         \end{minipage} $\to$
        \begin{minipage}[h]{0.2\linewidth}
	\vspace{0pt}
        \includegraphics[width=1\linewidth]{pic/zigzag2.png}
	 \end{minipage}
\caption{Examples of reduction of higher loop supergraphs to scalar conformal zig-zag master-integrals.}
\label{fig:ladder1}
\end{figure}

The other left $1/N$-subleading diagrams also have some kind of recurrence. The master-integrals for $1/N$-suppressed finite superdiagrams (see Fig.\ref{Fig:twoloopchiral}) are expected to be equivalent to the conformal zig-zags (see Fig.\ref{fig:ladder1})  or Broadhurst-Kreimer sequence with the corresponding normalization factor. These diagrams are known to be calculated exactly \cite{Broadhurst:1995km}. Recall that the Broadhurst-Kreimer conjecture for conformal zig-zags reads  \cite{Broadhurst:1995km,Derkachov:2022ytx}
\beq
\begin{gathered}
Z(L+1)= 4 C_L
\sum\limits_{p=1}^{\infty}
\frac{(-1)^{(p-1)(L+1)}}{p^{2(L+1)-3}} =
\left\{
\begin{array}{l}
4 \, C_L \, \zeta(2L-1) \;\; {\rm for} \;\; L =2N+1 \,,\\
4\, C_L \, (1 - 2^{2(1-L)}) \, \zeta(2L-1) \;\;
{\rm for} \;\; L=2N \,,
\end{array}
\right.
\end{gathered} \label{BK}
\eeq
where $L$ is a number of loops, $C_L = \frac{1}{(L+1)} \binom{2L}{L}$ is the Catalan number. It should be noted that the first nontrivial terms $6\zeta(3)$  to the zig-zag diagram were analytically evaluated in \cite{Chetyrkin:1980pr,Chetyrkin:1981qh}. The four-loop result is calculated in Refs.\cite{Kazakov:1983dyk,Kazakov:1983pk}.
Convenient operator representations for these diagrams in arbitrary order were obtained in the context of biscalar fishnet models in Refs.\cite{Derkachov:2022ytx,Derkachov:2023xqq}.

Thus, the whole finite subleading correction to the chiral effective superpotential can be expressed as
\begin{equation}    \textbf{W}^{sub} \sim g^{2L}~c_{L}/N^L\times Z(L+1) \times W_{tree},
\end{equation}
where $c_{L}$ is a sum of the symmetric factors of each supergraph of a given order.
It is quite possible that there is a general way to compute all $1/N$-subleading superdiagrams by perturbation theory and find a closed-form expression for the whole series at least under some assumptions or approximations, but we refer the study of this question for futher publications.

\section{Discussion}\label{sec:Concl}

We have performed one- and two-loop calculations of the chiral effective potential in the $\mathcal{N}=1$ supersymmetric gauge theory with chiral matter. The theory was formulated in terms of $\mathcal{N}=1$ superfields and investigated in the framework of the corresponding supergraph technique for the effective action depending on chiral superfields. One-loop chiral quantum corrections turn to be finite. As for two-loop corrections, it was shown that there are only a few types of two-loop supergraph topologies leading to a chiral effective superpotential.
The computation of the contributions of these supergraphs was given in Section \ref{sec:2loop} using the Wolfram Mathematica packages (e.g., \texttt{SusyMath.m} \cite{Ferrari:2007sc}) and methods for computing multiloop scalar master-integrals.

The computed diagrams are expectedly divided into divergent and finite diagrams. All of them are proportional to the tree-level superpotential.
At the one-loop level there is only one finite contribution to the chiral potential. The divergent two-loop diagrams include the one-loop two-point corrections to the vector and the chiral superfield Green function. In the case of finite subgraphs, the final two-loop contribution is finite. In the other case, the divergent contributions can be reduced by introducing appropriate counterterms at the diagram level or by renormalizing the fields in the original Lagrangian (see, e.g., \cite{Buchbinder:2025dgu} in case of two-loop chiral effective potential). 
The resulting one-loop contributions to the beta-function have been known for quite some time in the literature and are given in the references of this paper, e.g. \cite{WessBook:1992cp,Howe:1983wj,Parkes:1984dh,Storey:1981ee}. Otherwise, in the $\mathcal{N}=2$ model to which the discussed model can be tuned, one can find expressions for the beta-function and consider it as finiteness conditions. After satisfying the finiteness requirement only specific finite diagrams are left.

The finite two-loop superfield diagrams that give rise to the chiral effective potential have not been studied anywhere before and are given in this paper for the first time. Some sequences of the diagrams can be reduced to the conformal zig-zag diagrams, while in the planar limit, for some models we obtained the exact sum of all contributions to the chiral effective potential and found the result exact in coupling constant. In the considered finite $\mathcal{N}=2$ $SU(N)$ non-Abelian Yang-Mills model, summation of specific individual sequences of diagrams have been shown in the large $N$ limit.

In this context, it is interesting to find the connections of chiral contributions with conformal zig-zags and the possibility of obtaining exact values for chiral superpotentials in various models. Models in which finite contributions like those discussed above are also present and presumably computable using the superfield approach (like the  four-dimensional theories with deformations of $\mathcal{N}=4$ supersymmetry \cite{Buchbinder:1997ib,Fokken:2014soa,Kade:2024ucz}, as well as three-dimensional theories with extended supersymmetry \cite{Spiridonov:1991ki,Buchbinder:2012zd,Kade:2024lkc}) are particularly interesting for further study. Also, it would be interesting to develop methods for calculating chiral quantum corrections to the effective action in the gauge superfield sector in the manifestly $\mathcal{N}=1$ supersymmetric terms.

\subsection*{Acknowledgements}
The authors are grateful to I.B. Samsonov for useful comments. A. Mukhaeva's work is supported by the Foundation for the Advancement of Theoretical Physics and Mathematics BASIS, No 24-1-4-36-1.

\appendix

\section{D-algebra relations}\label{sec:AppA}

We introduce  the $2\times 2$ four-dimensional Pauli $\sigma$ matrices
\begin{equation}
\sigma^\mu=(\sigma_0,-\vec{\boldsymbol{\sigma}})\qquad\text{and}\qquad
\bar{\sigma}^\mu=(\sigma_0,\vec{\boldsymbol{\sigma}})\,,
\end{equation}
where $\vec{\boldsymbol{\sigma}}$ are the \textit{Pauli matrices}. We use the standard convention for raising/lowering the two-component spinor indices \(\alpha\), \(\dot \alpha\)
and introducing the tensors \(\varepsilon\) as the following relations hold:
\begin{align}
& (\bar \sigma^{\mu})^{\dot \alpha \alpha}= \varepsilon^{\dot\alpha \dot \beta} \varepsilon^{\alpha \beta} \sigma^{\mu}_{\dot \beta \beta},\qquad \varepsilon^{\dot\alpha \dot \beta}\varepsilon_{\dot \beta \dot \gamma}=\delta^{\dot \alpha}_{\dot \gamma}\qquad \varepsilon^{\alpha \beta}\varepsilon_{\beta \gamma}=\delta^{ \alpha}_{\gamma}\, .
\end{align}
The $\sigma$ matrices satisfy
\begin{equation}\label{symferm}
(\bar{\sigma}_\mu\sigma_\nu +\bar{\sigma}_\nu \sigma_\mu)^{\alpha}_\beta=-2\eta_{\mu\nu} \delta^\alpha_\beta \qquad\text{and}\qquad
(\sigma_\mu\bar{\sigma}_\nu +\sigma_\nu \bar{\sigma}_\mu)^{\dot{\alpha}}_{\dot{\beta}}=-2\eta_{\mu\nu} \delta^{\dot{\alpha}}_{\dot{\beta}} \,,
\end{equation}
and the trace identities are
\begin{equation}\begin{gathered}\label{tracesigma}
\text{tr}(\text{\textit{odd number of $\sigma$'s}})= 0,\\
\text{tr}(\sigma^\mu\bar{\sigma}^\nu)=\text{tr}(\bar{\sigma}^\mu\sigma^\nu)=-2\eta^{\mu\nu}.
\end{gathered}\end{equation}
Other complicated relations can be obtained using the identities mentioned above.
We also define the spinor covariant derivatives as
\beq
D_\alpha=\partial_\alpha+i (\sigma^\mu)_{\alpha \dot{\alpha}} \bar{\theta}^{\dot{\alpha}}\partial_\mu, ~\bar{D}_{\dot{\alpha}}=\partial_\alpha-i (\sigma^\mu)_{\alpha \dot{\alpha}} \theta^{\alpha}\partial_\mu,
\eeq
with the properties for the commutation relations to fulfill the $D$-algebra routine 
\begin{equation}
\begin{gathered}
\{D_\alpha,D_\beta\}=0,~ \{D_\alpha,\bar{D}_{\dot{\beta}}\}=-2i (\sigma)^\mu_{\alpha \dot{\beta}}\partial_\mu\\
D^2 \bar{D}_{\dot{\alpha}} D^2=0, \quad \bar{D}^2 D_\alpha \bar{D}^2=0, \\
D^\alpha \bar{D}^2 D_\alpha=\bar{D}_{\dot{\alpha}} D^2 \bar{D}^{\dot{\alpha}}, \\
D^2 \bar{D}^2+\bar{D}^2 D^2-2 D^\alpha \bar{D}^2 D_\alpha=16 \square \\
D^2 \bar{D}^2 D^2=16 \square D^2, \quad \bar{D}^2 D^2 \bar{D}^2=16 \square \bar{D}^2, \\
{\left[D^2, \bar{D}_{\dot{\alpha}}\right]=-4 i \partial_{\alpha \dot{\alpha}} D^\alpha, \quad\left[\bar{D}^2, D_\alpha\right]=4 i \partial_{\alpha \dot{\alpha}} \bar{D}^{\dot{\alpha}} .}
\end{gathered}
\end{equation}
The following identities are also valid for the integration measure
\beq
\begin{gathered}
    \int d^2\theta=-\frac{1}{4}D^2,~ \int d^2\bar{\theta}=\frac{1}{4}\bar{D}^2 \\
    d^2\theta=\frac{1}{4}\varepsilon^{\alpha \beta}d\theta_\alpha d\theta_\beta,~\int d\theta_\alpha \theta^\beta=\delta^\beta_\alpha.
\end{gathered}
\eeq
All the numerical factors for the supergraph vertices and integration measures are the same as in Ref. \cite{BK:book}.

\section{Feynman rules}\label{sec:Feyn_diagr}
\begin{figure}[h]
    \centering
    \includegraphics[width=1\linewidth]{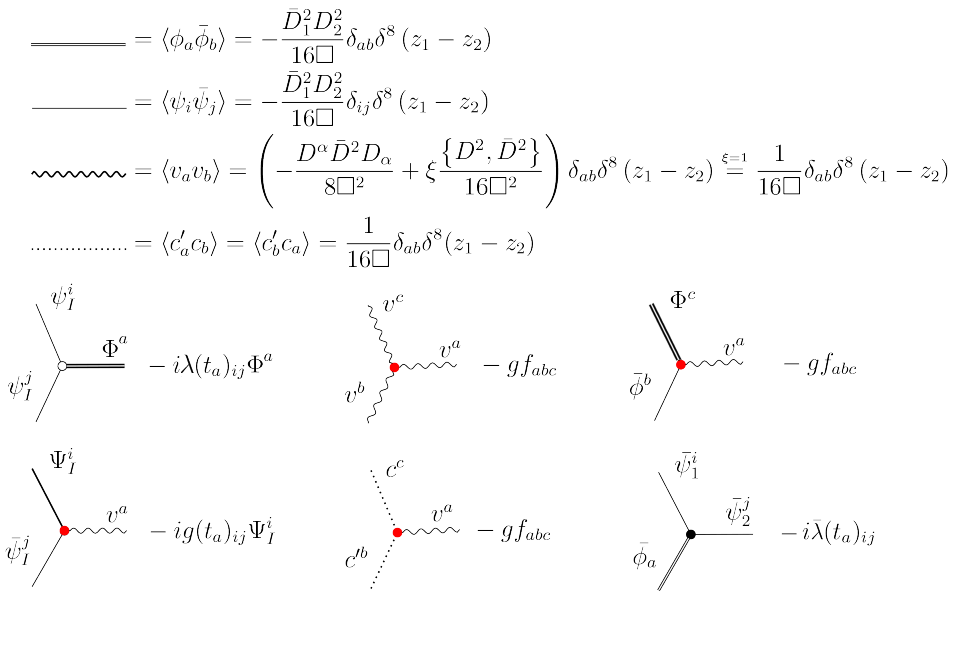}
    \caption{Feynman rules necessary for calculating chiral contributions in the models considered in the main text}
    \label{fig:Feynrules}
\end{figure}

The quadratic superpropagator part of the effective action from the conventions and discussion above, which is needed to calculate the chiral effective superpotential, can be obtained in the form presented in \cite{BK:book,Grisaru:1996ve,Grisaru:SUPERGRAPHITY}. In order to study the chiral potential, Green functions should not have antichiral external lines. This requirement is equivalent to a restriction on the number of covariant and anticovariant $D$-derivatives mentioned in the Introduction. Keeping this statement in mind, one can obtain Feynman rules for the chiral effective superpotential\footnote{In principle, one could derive all the same rules by decomposing the action in the external field and reexpressing and decomposing the Green functions over the chiral superfield, as was done, for example, in \cite{BKY:1994iw,Buchbinder:2025dgu}};
so the propagators and vertices can be found in the form depicted in Fig. \ref{fig:Feynrules}.
Here one should also take into account the corresponding conjugates of the depicted vertices  \cite{BK:book,Grisaru:Improved,Grisaru:SUPERGRAPHITY,WestBook}.

\section{Scalar master-integrals}\label{sec:AppB}
\subsection{Two-point functions}
The one-loop two-point integral is given by the
\beq
J^{(1)}_{\alpha,\beta}(k)=\int \frac{d^d q}{(q-k)^{2\alpha}q^{2\beta}}=\frac{a(\alpha)a(\beta)}{a(\alpha+\beta-d/2)}(k^2/\mu^2)^{d/2-\alpha-\beta},
\label{bubble}
\eeq
where $a(\alpha)=\Gamma(d/2-\alpha)/\Gamma(\alpha)$. For instance,
\beq
J^{(1)}_{1,1}(k)=(k^2/\mu^2)^{-\epsilon}\left(\frac{1}{\epsilon}+2+O(\epsilon^1)\right).
\label{eq:J_11}
\eeq
To calculate more complicated diagrams, e.g., with numerators, one can refer to \cite{Kotikov:2018wxe}.

Another useful diagram is the walnut-type diagram. This two-loop three-point integral at zero external momenta is given by the
\beq
\begin{gathered}
J^{(2)}=\lim_{p_1,p_2\rightarrow0}\int \frac{d^4 q_1}{(4\pi)^2}\frac{d^4q_2}{(4\pi)^2} \frac{q_1^2 (p_1+p_2)^2}{q_1^2 (q_1-p_1)^2 (q_2-p_2)^2 q_2^2 (q_2-p_2)^2  (q_1-p_2)^2}=6\zeta(3).
\end{gathered}
\eeq
The contributions  of the third-loop and fourth-loop can be obtained using the uniqueness method, but here we present only the result \cite{Kazakov:1983pk,Chetyrkin:1980pr}
\beq
\begin{gathered}
J^{(3)}=20\zeta(5),~J^{(4)}=\frac{441}{8}\zeta(7),
\end{gathered}
\eeq
etc., in accordance with \eqref{BK}.
Higher loops of these integrals can be reduced to integrals obtained iteratively by means of the one-magnon graph-building operator in the conformal $4D$ biscalar fishnet model, as shown in Refs. \cite{Derkachov:2022ytx,Derkachov:2023xqq}.
\subsection{Three-point functions}
The Usyukina-Davydychev triangle integral from
\cite{Usyukina:1992jd,Usyukina:1993ch} for $d=4-2\epsilon$ can be represented as expansion over small $\epsilon$:
\beq
\label{expan}
\Upsilon_L(u,v) = \frac{\Gamma(1-\epsilon)}{u^{2(1-\epsilon)} }
\sum_{l=0}^{\infty} \frac{\epsilon^l}{l!} \, \Upsilon^{(l)}_L(z_1,z_2) \; .
\eeq
Here the dimensionless parameters $z_1,z_2$ are defined
by the equations: $z_1 + z_2 = 2(uv)/u^2$ and $z_1z_2 = v^2/u^2$, and $u,v$ are the conformal kinematic variables in the notation of \cite{Usyukina:1992jd,Isaev:2003tk}. The functions $\Upsilon^{(l)}_L$ in the expansion (\ref{expan}) can be represented in the following form:
\beq
\label{coef}
\Upsilon^{(l)}_L(z_1,z_2) =
 \sum_{f=0}^{L} \, \frac{(-\ln(z_1z_2))^f \, (2L-f)}{ f! \, (L-f)!}
 \sum_{m=0}^l   \frac{(-1)^m ~l!}{m! (l-m)!}
 {\bf Z}_{m}\left(z_1,z_2;2L +l-f \right) \; ,
\eeq
 and we denote $(k > m)$:
 $$
 {\bf Z}_{m}(z_1,z_2;k) =
\frac{\Gamma(k-m)}{(z_1 -z_2) } \; \sum_{\{n_i\} =1}^{\infty}
\frac{(z_1^{n_0} - z_2^{n_0})}{\left( \sum_{0}^m n_i \right)^{k-m} }
 \left(\prod^m_{i=1} \frac{z_1^{n_i} + z_2^{n_i}}{n_i} \right) \;,
$$
at  zero external momenta $u,v \rightarrow1$ for $l=0$ (as we are interested only in finite contributions to the chiral effective potential) these expressions are considerably simplified
\beq
\begin{gathered}
\Upsilon^{(l)}=\int_0^1 d\tau \frac{2 \log^{2l-1}(\tau)}{\tau^2-\tau+1}=\\=\frac{1}{2^{2l-1} 3^{2l}}  \left(\psi ^{(2l+1)}\left(\frac{2}{3}\right)-\psi ^{(2l+1)}\left(\frac{1}{3}\right)-\psi ^{(2l+1)}\left(\frac{1}{6}\right)+\psi ^{(2l+1)}\left(\frac{5}{6}\right)\right),
\label{eq:master1}
\end{gathered}
\eeq
where $\psi^{(n)}$ is the polygamma function.
Further simplifications of these functions can be carried out in terms of the multiple zeta or Clausen functions; however, for the purposes of this work, the result in this analytical form is sufficient.

\newpage

\bibliographystyle{JHEP}
\bibliography{refs}
\end{document}